\documentclass{cmspaper}

\newcommand{\G}{\gamma}
\newcommand{\GG}{{\gamma\gamma}}
\newcommand{\EPEM}{e^+e^-}
\newcommand{\MUPM}{\mu^+\mu^-}
\newcommand{\A}{\alpha}
\newcommand{\ZA}{Z\alpha}
\newcommand{\BE}{\begin{equation}}
\newcommand{\EE}{\end{equation}}

\makeatletter
\def\lesssim{\mathrel{\mathpalette\vereq<}}
\def\vereq#1#2{\lower3pt\vbox{\baselineskip1.5pt \lineskip1.5pt
\ialign{$\m@th#1\hfill##\hfil$\crcr#2\crcr\sim\crcr}}}
\def\gtrsim{\mathrel{\mathpalette\vereq>}}
\makeatother

\begin{document}

\begin{titlepage}
\cmsnote{1999/xxx}
\date{Jul 23, 1999}

\title{Coherent Interactions with heavy ions at CMS}

\begin{Authlist}
G.~Baur
 \Instfoot{fzj}{Forschungszentrum J\"ulich, J\"ulich, Germany}
K.~Hencken, D.~Trautmann
 \Instfoot{uba}{Universit\"at Basel, Basel, Switzerland}
S.~Sadovsky and Yu.~Kharlov
 \Instfoot{ihep}{IHEP, Protvino, Russia}
\end{Authlist}


\note{Version 1.3}

\end{titlepage}

\setcounter{page}{2}

\section*{General Introduction}
\label{sec_introduction}

The physics of central collisions is the physics of the Quark Gluon
Plasma. Apart from projects like the search for new physics at very high 
rapidities (see the CASTOR subproject at ALICE for a search
for Centauro events at LHC \cite{Angelis99}),
``Non QGP Physics'' may be defined as the physics of
peripheral collisions, which includes the effects of coherent photons
and diffraction effects (Pomeron exchange). It is our aim to show that
one will be able at CMS to address very interesting physics topics in a
rather clean way.

Central collision events are characterized by a very high
multiplicity. On the other hand, the multiplicity in peripheral
collisions is comparatively low. The ions do not interact directly 
with each other and move on essentially undisturbed in the beam
direction. The only possible interaction are therefore due to
the long range electromagnetic interaction and diffractive processes.
Due to the coherent action of all the protons
in the nucleus, the electromagnetic field is very strong and the
resulting flux of equivalent photons is large. It is proportional to
$Z^2$, where Z is the nuclear charge. Due to the very short
interaction times the spectrum of these photons extends up to an
energy of about 100GeV in the laboratory system. The coherence
conditions limits the virtuality of the photon to very low values of
$Q^2 < 1/R^2$, where $R=1.2fm A^{1/3}$ is the nuclear size.

%
Hard diffractive processes in heavy ion collisions have also been studied.
These are interesting processes on their own, but they are
also a possible background to photon-photon and photon-hadron
interactions. The physics potential of such kind of collisions is
discussed in Section \ref{sec_photon} (this is an extension of CMS
note1998/009). It ranges from studies in QCD and strong field QED to
the search for new particles (like a light Higgs particle). This kind
of physics is strongly related to $\GG$ physics at
$\EPEM$-colliders with increased luminosity. In view of the strong
interaction background, experimental conditions will be somewhat
different from the $\GG$ physics at $\EPEM$-colliders. A
limitation of the heavy ions is that only quasireal but no highly
virtual photons will be available in the A-A collisions. 

Another aspect is the study of photon-hadron interactions, extending
the $\G$-p interaction studies at HERA/DESY to $\G$-A interactions,
also reaching higher invariant masses than those possible at HERA.  

At the STAR (Solenoidal Tracker At RHIC) detector at RHIC --- to be
scheduled to begin taking data in 1999 --- a program to study
photon-photon and Pomeron interactions in peripheral collisions
exists \cite{KleinS97a,KleinS97b,KleinS95a,KleinS95b,Nystrand98}. 
At RHIC the photon flux will be of the same order of
magnitude, but the spectrum is limited to up to about 3 GeV.

\section{Photon-Photon and Photon-Hadron Physics}
\label{sec_photon}

\subsection{Abstract}
\label{ssec_abstract}

Due to coherence, there are strong electromagnetic
fields of short duration in very peripheral collisions. They give rise to
photon-photon and photon-nucleus collisions with high flux up to an
invariant mass region hitherto unexplored experimentally. After a
general survey photon-photon luminosities in relativistic heavy ion
collisions are discussed. Special care is taken to include the effects
of strong interactions and nuclear size. Then photon-photon physics at
various $\GG$-invariant mass scales is discussed.
Invariant masses of up to about 100
GeV can be reached at LHC, and in addition the potential for new
physics is available. Photonuclear reactions and other important background
effects, mainly diffractive processes are also discussed.
Lepton-pair production, especially electron-positron pair production
is copious. Due to the strong fields there will be new phenomena,
like multiple $\EPEM$ pair production.

\subsection{Introduction}
\label{ssec_intro}

The parton model is very useful to study scattering processes at very
high energies. The scattering is described as an incoherent
superposition of the scattering of the various constituents. For
example, nuclei consist of nucleons which in turn consist of quarks
and gluons, photons consist of lepton pairs, electrons consist of
photons, etc.. We note that relativistic nuclei have photons as an
important constituent, especially for low enough virtuality
$Q^2=-q^2>0$ of the photon. This is due to the coherent action of all
the charges in the nucleus.  The virtuality of the photon is related
to the size $R$ of the nucleus by
\BE
Q^2 \lesssim 1/R^2, 
\EE
the condition for coherence. The radius of a nucleus is given
approximately by $R=1.2$~fm~$A^{1/3}$, where $A$ is the nucleon
number. From the kinematics of the process one has
\BE
Q^2=\frac{\omega^2}{\G^2}+q_\perp^2,
\EE
where $\omega$ and $q_\perp$ are energy and transverse momentum of the
quasireal photon. This limits the maximum energy of the quasireal
photon to 
\BE
\omega<\omega_{max} \approx \frac{\G}{R},
\label{eq_wmax}
\EE
where $\G$ is the Lorentz factor of the projectile
and the perpendicular component of its momentum to
\BE
q_\perp \lesssim \frac{1}{R}.
\EE
We define the
ratio $x=\omega/E$, where $E$ denotes
the energy of the nucleus $E= M_N \G A$ and $M_N$ is the nucleon mass.
It is therefore smaller than
\BE
x< x_{max}=\frac{1}{R M_N A} = \frac{\lambda_C(A)}{R},
\EE
where $\lambda_C(A)$ is the Compton wave length of the ion. Here and
also throughout the rest of the paper we use natural units, setting
$\hbar=c=1$.
%
%
\begin{figure}[tbh]
\begin{center}
\resizebox{4.5cm}{!}{\includegraphics{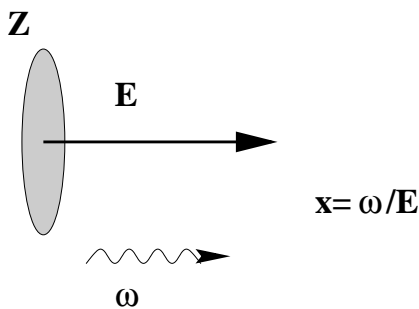}}
\end{center}
\caption{\it
A fast moving nucleus with charge $Ze$ is surrounded by a strong
electromagnetic field. This can be viewed as a cloud of virtual
photons.  These photons can often be considered as real. They are
called equivalent or quasireal photons. The ratio of the photon energy
$\omega$ and the incident ion energy $E$ is denoted by $x=\omega/E$.
Its maximal value is restricted by the coherence condition to
$x<\lambda_C(A)/R\approx 0.175/A^{4/3}$, that is, $x\protect\lesssim
10^{-3}$ for Ca ions and $x\protect\lesssim 10^{-4}$ for Pb ions.  }
\label{fig_xvar}
\end{figure}

The collisions of $e^+$ and $e^-$ has been the traditional way to
study $\GG$-collisions. Similarly photon-photon collisions can also be
observed in hadron-hadron collisions. Since the photon number scales
with $Z^2$ ($Z$ being the charge number of the nucleus) such effects
can be particularly large. Of course, the strong interaction of the
two nuclei has to be taken into consideration.
%
%
\begin{figure}[tbh]
\begin{center}
\resizebox{5cm}{!}{\includegraphics{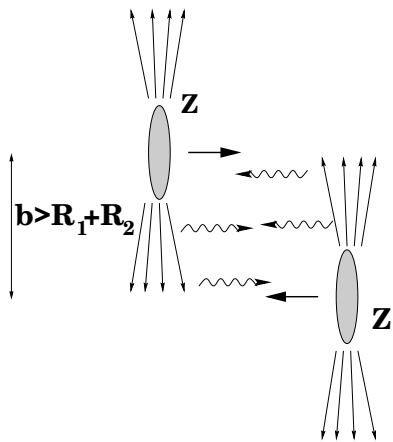}}
\end{center}
\label{fig_collision}
\caption{\it
Two fast moving electrically charged objects are an abundant source of
(quasireal) photons. They can collide with each other and with the
other nucleus. For peripheral collisions with impact parameters $b>2R$,
this is useful for photon-photon as well as photon-nucleus
collisions.}
\end{figure}

The equivalent photon flux present in medium and high energy nuclear
collisions is very high. Recent reviews of the present topic
can be found in \cite{BaurHT98,KraussGS97,BaurHT98b,HenckenSTB99}.
This high equivalent photon flux has already
 found many useful applications in
nuclear physics \cite{BertulaniB88}, nuclear astrophysics
\cite{BaurR94,BaurR96}, particle physics \cite{Primakoff51} (sometimes
called the ``Primakoff effect''), 
as well as, atomic physics \cite{Moshammer97}.
Here our main purpose is to discuss the physics of photon-photon and
photon-hadron (nucleus) collisions in high energy heavy ion
collisions. The ``Relativistic Heavy Ion Collider'' (RHIC),
scheduled to begin taking data in 1999, will have a program to investigate
such collisions experimentally. The equivalent photon spectrum there
extends up to several GeV ($\G\approx 100$). Therefore the available
invariant mass range is up to about the mass of the $\eta_c$.  
At the recent RHIC/INT (``Institute for Nuclear Theory'') workshop at
the LBNL (Berkeley), the physics of peripheral collisions was discussed
by S. R. Klein and S. J. Brodsky \cite{rhicint99}. 
When the ``Large Hadron Collider'' will be scheduled to begin taking data
in 2004/2008, the study of these reactions can be extended to both
higher luminosities but also to much higher invariant masses, hithero
unexplored. 

Relativistic heavy ion collisions have been suggested as a general
tool for two photon physics about a decade ago. Yet the study of a
special case, the production of $\EPEM$ pairs in nucleus-nucleus
collisions, goes back to the work of Landau and Lifschitz in 1934
\cite{LandauL34} (In those days, of course, one thought more about
high energy cosmic ray nuclei than relativistic heavy ion
colliders).
The general possibilities and characteristic features of two-photon
physics in relativistic heavy ion collisions have been discussed in
\cite{BaurB88}. The possibility to produce a Higgs boson via
$\GG$-fusion was suggested in \cite{GrabiakMG89,Papageorgiu89}. In
these papers the effect of strong absorption in heavy ion collisions
was not taken into account. This absorption is a feature, which is
quite different from the two-photon physics at $\EPEM$ colliders. The
problem of taking strong interactions into account was solved by using
impact parameter space methods in \cite{Baur90d,BaurF90,CahnJ90}. Thus
the calculation of $\GG$-luminosities in heavy ion collisions is put
on a firm basis and rather definite conclusions were reached by many
groups working in the field, as described, e.g., in
\cite{VidovicGB93,KraussGS97,BaurHT98}. This opens the way
for many interesting applications.
Up to now hadron-hadron collisions have not been used for
two-photon physics. An exception can be found in
\cite{Vannucci80}, where the production of $\mu^+\mu^-$ pairs at the
ISR was studied.  The special class of events was selected, where no
hadrons are seen associated with the muon pair in a large solid angle
vertex detector. In this way one makes sure that the hadrons do not
interact strongly with each other, i.e., one is dealing with
peripheral collisions (with impact parameters $b>2R$); the
photon-photon collisions manifest themselves as ``silent events'',
that is, with only a small relatively small multiplicity.
Dimuons with a very low sum of transverse momenta are also considered
as a luminosity monitor for the ATLAS detector at LHC \cite{ShamovT98}.

Experiments are planned at RHIC
\cite{KleinS97a,KleinS97b,KleinS95a,KleinS95b,Nystrand98} and are
discussed at LHC
\cite{HenckenKKS96,Sadovsky93,BaurHTS98}.
We quote J. D. Bjorken \cite{Bjorken99}:
{\it It is an important portion (of the FELIX program at LHC
\cite{Felix97}) to tag on
Weizsaecker Williams photons (via the nonobservation of completely
undissociated forward ions) in ion-ion running, creating a high
luminosity $\GG$ collider.}

\subsection{From impact-parameter dependent equivalent photon spectra to
{$\GG$-luminosities}}
\label{ssec_lum}

\begin{figure}[tbhp]
\begin{center}
\resizebox{3cm}{!}{\includegraphics{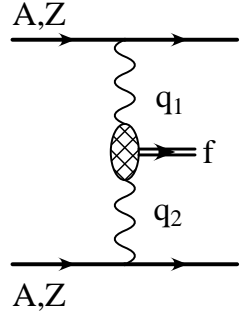}}
\end{center}
\caption{\it
The general Feynman diagram of photon-photon processes in heavy ion
collisions: Two (virtual) photons fuse in a charged particle collision
into a final system $f$.
} 
\label{fig_ggcollision}
\end{figure}

Photon-photon collisions have been studied extensively at $\EPEM$
colliders. The theoretical framework is reviewed, e.g., in
\cite{BudnevGM75}.  The basic graph for the two-photon process in
ion-ion collisions is shown in Fig.~\ref{fig_ggcollision}. Two virtual
(space-like) photons collide to form a final state $f$. In the
equivalent photon approximation (EPA) it is assumed that the square of the
4-momentum of the virtual photons is small, i.e., $q_1^2\approx
q_2^2\approx 0$ and the photons can be treated as quasireal. In this
case the $\GG$-production is factorized into an elementary cross
section for the process $\G+\G\rightarrow f$ (with real photons, i.e.,
$q^2=0$) and a $\GG$-luminosity function. In contrast to the pointlike
elementary electrons (positrons), nuclei are extended, strongly
interacting objects with internal structure. This gives rise to
modifications in the theoretical treatment of two photon processes.
The emission of a photon depends on the (elastic) form factor. Often a
Gaussian form factor or one of a homogeneous charged sphere is used.
The typical behavior of a form factor is
\BE
f(q^2) \approx 
\left\{
 \begin{array}{lcl}
 Z &\qquad& \mbox{for $|q^2| < \frac{1}{R^2}$}\\
 0 &\qquad& \mbox{for $|q^2| \gg \frac{1}{R^2}$}
 \end{array}
\right. .
\EE
For low $|q^2|$ all the protons inside the nucleus act coherently,
whereas for $|q^2| \gg 1/R^2$ the form factor is very small, close to
0. For a medium size nucleus with, say, $R=5$ fm, the limiting
$Q^2=-q^2=1/R^2$ is given by $Q^2=(40$MeV$)^2=1.6\times
10^{-3}$~GeV${}^2$. Apart from $\EPEM$ (and to a certain extent also
$\mu^+\mu^-$) pair production, this scale is much smaller than typical
scales in the two-photon processes. Therefore the virtual photons in
relativistic heavy ion collisions can be treated as quasireal. This is
a limitation as compared to $\EPEM$ collisions, where the two-photon
processes can also be studied as a function of the corresponding
masses $q_1^2$ and $q_2^2$ of the exchanged photon (``tagged mode'').
%
%

As was discussed already in the previous section, relativistic
heavy ions interact strongly when the impact parameter is smaller than
the sum of the radii of the two nuclei. In such cases $\GG$-processes
are still present and are a background that has to be considered in
central collisions. In order to study ``clean'' photon-photon events
however, they have to be eliminated in the calculation of
photon-photon luminosities as the particle production due to the
strong interaction dominates. In the usual treatment of photon-photon
processes in $\EPEM$ collisions plane waves are used and there is
no direct information on the impact parameter. For heavy ion
collisions on the other hand it is very appropriate to introduce
impact parameter dependent equivalent photon numbers. They have been
widely discussed in the literature, see, e.g.,
\cite{BertulaniB88,JacksonED,WintherA79}.

The equivalent photon spectrum corresponding to a point charge $Z e$,
moving with a velocity $v$ at impact parameter $b$ is given by
\BE
N(\omega,b) = \frac{Z^2\A}{\pi^2} \frac{1}{b^2}
\left(\frac{c}{v}\right)^2 x^2 \left[ K_1^2(x) + \frac{1}{\G^2}
K_0^2(x)\right],
\label{eq_nomegab}
\EE
where $K_n(x)$ are the modified Bessel Functions (MacDonald
Functions) and $x=\frac{\omega b}{\G v}$. 
Then one obtains the probability for a certain electromagnetic process
to occur in terms of the same process generated by an equivalent pulse
of light as
\BE
P(b) = \int \frac{d\omega}{\omega} N(\omega,b) \sigma_\G(\omega).
\EE
Possible modifications of $N(\omega,b)$ due to an extended spherically
symmetric charge distribution are given in \cite{BaurF91}. It should
be noted that Eq.~(\ref{eq_nomegab}) also describes the equivalent
photon spectrum of an extended charge distribution, such as a nucleus,
as long as $b$ is larger than the extension of the object. This is due
to the fact that the electric field of a spherically symmetric system
depends only on the total charge, which is inside it. 

As the term $x^2 \left[ K_1^2(x) + 1/\G^2 K_0^2(x)\right]$ in
Eq.~(\ref{eq_nomegab}) can be roughly approximated as 1 for $x<1$ and
0 for $x>1$, that is, the equivalent photon number
$N(\omega,b)$ is almost a constant up to a maximum
$\omega_{max}=\G/b$ ($x=1$). By integrating
the photon spectrum (Eq.~(\ref{eq_nomegab}))
over $b$ from a minimum value of $R_{min}$ up to infinity (where
essentially only impact parameter up to  $b_{max}\approx \G/\omega$ 
contribute, compare with Eq.~(\ref{eq_wmax})),
one can define an equivalent photon number
$n(\omega)$. This integral can be carried out analytically and is
given by \cite{BertulaniB88,JacksonED}
\BE
n(\omega) = \int d^2b N(\omega,b) = \frac{2}{\pi} Z_1^2 \alpha
\left(\frac{c}{v}\right)^2 \left[ \xi K_0 K_1 - \frac{v^2\xi^2}{2 c^2}
\left(K_1^2 - K_0^2\right)\right] ,
\label{eq_nomegaex}
\EE
where the argument of the modified Bessel functions is 
$\xi=\frac{\omega R_{min}}{\G v}$.
The cross section for a certain electromagnetic process is then
\BE
\sigma = \int \frac{d\omega}{\omega} n(\omega) \sigma_{\G}(\omega).
\label{eq_sigmac}
\EE
Using the approximation above for the MacDonald functions, we get
an approximated form, which is quite reasonable and is useful for
estimates:
\BE
n(\omega) \approx  \frac{2 Z^2 \A}{\pi} \ln
\frac{\G}{\omega R_{min}} \qquad \omega<\gamma/R_{min}.
\label{eq_nomegaapprox}
\EE

The photon-photon production cross-section is obtained in a similar
factorized form,
by folding the corresponding equivalent photon spectra of the two
colliding heavy ions \cite{BaurF90,CahnJ90} (for
polarization effects see \cite{BaurF90}, they are neglected here)
\BE
\sigma_c = \int \frac{d\omega_1}{\omega_1} \int
\frac{d\omega_2}{\omega_2}
F(\omega_1,\omega_2) \sigma_{\GG}(W_{\GG}) ,
\label{eq_sigmaAA}
\EE
with
\begin{eqnarray}
F(\omega_1,\omega_2)&=& 2\pi \int_{R_1}^{\infty} b_1 db_1 
\int_{R_2}^{\infty} b_2 db_2 \int_0^{2\pi} d\phi \nonumber\\
&&\times N(\omega_1,b_1) N(\omega_2,b_2)
\Theta\left(b_1^2+b_2^2-2 b_1 b_2
\cos\phi-R_{cutoff}^2\right) ,
\label{eq_fw1w2}
\end{eqnarray}
where $W_{\GG}=\sqrt{4 \omega_1\omega_2}$ is the invariant mass of the 
$\GG$-system and $R_{cutoff} = R_1 + R_2$.  
(In \cite{Nystrand98} the effect of replacing the simple sharp cutoff
($\Theta$-function) by a more realistic probability of the nucleus to
survive is studied. Apart from the very high end of the spectrum,
modifications are rather small.)
This can also be rewritten in terms of
the invariant mass $W_{\GG}$ and the rapidity
$Y=1/2 \ln[(P_0+P_z)/(P_0-P_z)]=1/2 \ln(\omega_1/\omega_2)$ as:
\BE
\sigma_c = \int dW_{\GG} dY \frac{d^2L}{dW_{\GG} dY}
\sigma_{\GG}(W_{\GG}) ,
\label{eq_sigmaAAMY}
\EE
with 
\BE
\frac{d^2L_{\GG}}{dW_{\GG} dY} = \frac{2}{W_{\GG}}
F\left(\frac{W_{\GG}}{2} e^Y,\frac{W_{\GG}}{2} e^{-Y}\right) .
\label{eq_dldwdy}
\EE
Here energy and momentum of the $\GG$-system in the beam direction are
denoted by $P_0$ and $P_z$. The transverse momentum is of the order of
$P_\perp \le 1/R$ and is neglected here. The transverse momentum
distribution is calculated in \cite{BaurB93}.

In \cite{BaurB93} and \cite{Baur92}
the intuitively plausible formula Eq.~(\ref{eq_fw1w2}) is derived ab
initio, starting from the assumption that the two ions move on a
straight line with impact parameter $b$. 
The advantage of heavy nuclei is seen in the coherence factor $Z_1^2
Z_2^2$ contained in the $N(\omega,b)$ in Eq.~(\ref{eq_fw1w2}).

\begin{figure}[tbh]
\begin{center}
\resizebox{8cm}{!}{\includegraphics{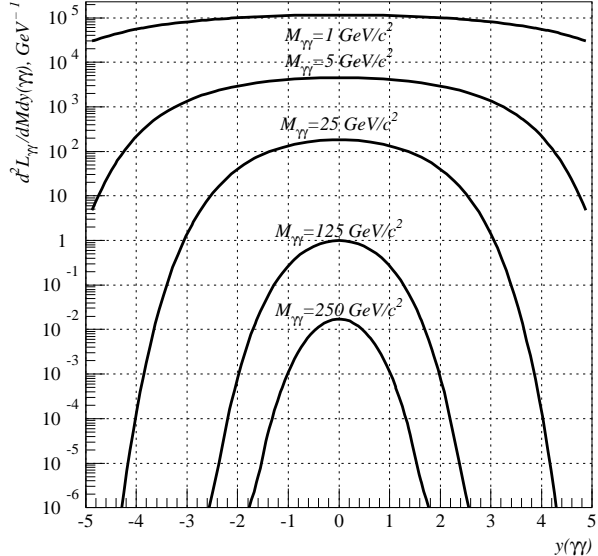}}
\end{center}
\caption{
The luminosity function $d^2L_{\GG}/dMdY$ for Pb-Pb collisions with 
$\gamma=2950$ as a function of $Y$ for different values of $M$.
}
\label{fig_y}
\end{figure}

As a function of $Y$, the luminosity $d^2L/dW_{\GG}{dY}$ for symmetrical
ion collisions has a
Gaussian shape with the maximum at $Y=0$. The width is approximately
given by $\Delta Y = 2 \ln \left[(2\G)/(R W_{\GG})\right]$, see also 
Fig.~\ref{fig_y}. Depending on the experimental situation additional cuts 
in the allowed $Y$ range are needed.

Additional effects due to the nuclear structure have been also studied.
For inelastic vertices a photon number $N(\omega,b)$ can also be defined, 
see, e.g., \cite{BaurHT98}. Its effect was found to be small. The
dominant correction comes from the electromagnetic excitation of one
of the ions in addition to the photon emission. 
We refer to \cite{BaurHT98} for further details.

\begin{figure}[tbh]
\begin{center}
\resizebox{8cm}{!}{\includegraphics{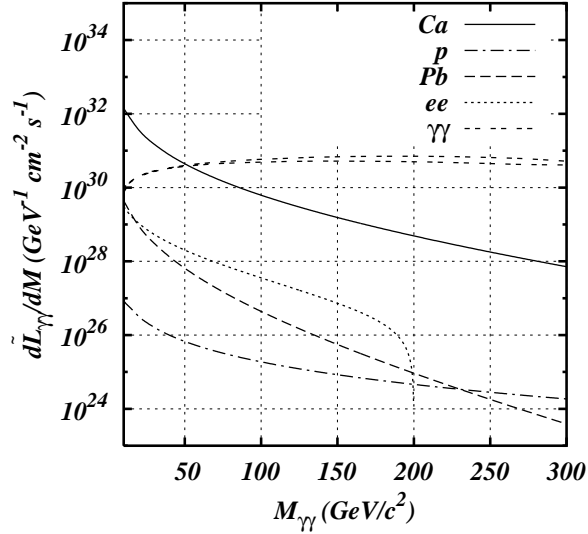}}
\end{center}
\caption{\it Comparison of the effective $\GG$-Luminosities 
($d\tilde L_{\GG}/dM = L_{AA} \times dL_{\GG}/dM$) for different ion 
species. For comparison the same
quantity is shown for LEP200 and a future NLC/PLC (next linear
collider/photon linear collider), where photons are obtained by laser
backscattering; the results for two different polarizations are shown.}
\label{fig_lum}
\end{figure}

In Fig.~\ref{fig_lum}
we give a comparison of effective $\GG$ luminosities, that is the product
of the beam luminosity with the two-photon luminosity 
($L_AA \times dL_{\GG}/dM$) for various collider scenarios.
We use the following collider parameters: LEP200: $E_{el}=100$GeV,
$L=10^{32} cm^{-2} s^{-1}$, NLC/PLC: $E_{el}=500$GeV, 
$L=2 \times 10^{33} cm^{-2} s^{-1}$, Pb-Pb heavy-ion mode at LHC: 
$\gamma=2950$,
$L=10^{26} cm^{-2} s^{-1}$, Ca-Ca: $\gamma=3750$,
$L=4 \times 10^{30} cm^{-2} s^{-1}$,p-p: $\gamma=7450$,
$L=10^{30} cm^{-2} s^{-1}$.
In the Ca-Ca heavy ion mode, higher 
effective luminosities (defined as collider luminosity times
$\GG$-luminosity) can be achieved as, e.g., in the Pb-Pb mode, 
since higher AA luminosities can be reached there.
Since the event rates are proportional to the luminosities, and 
interesting events are rare (see also below), we think that it is important
to aim at rather high luminosities in the ion-ion runs.
This should be possible,especially for the medium heavy ions like Ca.
For further details see \cite{BrandtEM94,Bruening98,HenckenTB95}.

\subsection{$\G$-A interactions}
\label{ssec_ga}

There are many interesting phenomena ranging from the excitation of
discrete nuclear states, giant multipole resonances (especially the
giant dipole resonance), quasideuteron absorption, nucleon resonance
excitation to the nucleon continuum.

The interaction of quasireal photons with protons has been studied
extensively at the electron-proton collider HERA (DESY, Hamburg), with
$\sqrt{s} = 300$~GeV ($E_e=27.5$~GeV and $E_p=820$~GeV in the
laboratory system). This is made possible by the large flux of
quasi-real photons from the electron (positron) beam. The obtained $\G
p$ center-of-mass energies (up to $W_{\G p}\approx200$~GeV) are an
order of magnitude larger than those reached by fixed target
experiments. 

Similar and more detailed studies will be possible at the
relativistic heavy ion colliders RHIC and LHC, due to the larger flux
of quasireal photons from one of the colliding nuclei. In the
photon-nucleon subsystem, one can reach invariant masses $W_{\G N}$ up
to $W_{\G N,max}=\sqrt{4 W_{max} E_N} \approx 0.8 \G A^{-1/6}$~GeV.For
Pb at LHC ($\G=2950$) one obtains 950~GeV and even higher
values for Ca. Thus one can study
physics quite similar to the one at HERA, with nuclei instead of
protons. Photon-nucleon physics includes many aspects, like the energy
dependence of total cross-sections, diffractive and non-diffractive
processes.

An important subject is the elastic vector meson production $\G p
\rightarrow V p$ (with $V=\rho,\omega,\phi,J/\Psi,\dots$). A review of
exclusive neutral vector meson production is given in
\cite{Crittenden97}.  The diffractive production of vector mesons
allows one to get insight into the interface between perturbative QCD
and hadronic physics. Elastic processes (i.e., the proton remains in
the ground state) have to be described within nonperturbative (and
therefore phenomenological) models. It was shown in \cite{RyskinRML97}
that diffractive (``elastic'') $J/\Psi$ photoproduction is a probe of
the gluon density at $x\approx \frac{M_{\Psi}^2}{W_{\G N}^2}$ (for
quasireal photons).  Inelastic $J/\Psi$ photoproduction was also
studied recently at HERA \cite{Breitweg97}.

Going to the hard exclusive photoproduction of heavy mesons on the
other hand, perturbative QCD is applicable. Recent data from HERA on
the photoproduction of $J/\Psi$ mesons have shown a rapid increase of
the total cross section with $W_{\G N}$, as predicted by perturbative
QCD.  Such studies could be extended to photon-nucleus interactions at
RHIC, thus complementing the HERA studies. Equivalent photon flux
factors are large for the heavy ions due to coherence. On the other
hand, the A-A luminosities are quite low, as compared to HERA. Of
special interest is the coupling of the photon of one nucleus to the
Pomeron-field of the other nucleus. Such studies are envisaged for
RHIC, see \cite{KleinS97a,KleinS97b,KleinS95a,KleinS95b} where also
experimental feasibility studies were performed.

Estimates of the order of magnitude of vector meson production in
photon-nucleon processes at RHIC and LHC are given in \cite{BaurHT98}.  In
$AA$ collisions there is incoherent photoproduction on the individual
$A$ nucleons. Shadowing effects will occur in the nuclear environment
and it will be interesting to study these \cite{BauerSYP78}. There is
also the coherent contribution where the nucleus remains in the ground
state.  Due to the large momentum transfer, the total (angle
integrated) coherent scattering shows an $A^{4/3}$ dependence. (It
will be interesting to study shadow effects in this case also). This
is in contrast to, e.g., low energy $\nu$A elastic scattering, where
the coherence effect leads to an $A^2$ dependence.  
For a general pedagogical discussion of the coherence effects see, e.g., 
\cite{FreedmanST77}. 
The coherent exclusive vector meson production at RHIC was studied
recently in \cite{KleinN99}. The increase of the cross section
with $A$ was found there to be between the two extremes ($A^{4/3}$ and
$A^2$) mentioned above. In this context, RHIC and LHC can be considered 
as vector meson factories \cite{KleinN99}. In addition there are 
inelastic contributions, where the proton (nucleon) is transformed into 
some final state $X$ during the interaction (see \cite{Breitweg97}). 

At the LHC one can extend these processes to much higher invariant
masses $W$, therefore much smaller values of $x$ will be probed.
Whereas the $J/\Psi$ production at HERA was measured up to invariant
masses of $W\approx 160$~GeV, the energies at the LHC allow for
studies up to $\approx 1$~TeV.

At the LHC \cite{Felix97} hard diffractive vector
meson photoproduction can be investigated especially well in $AA$
collisions. In comparison to previous experiments, the very large
photon luminosity should allow observation of processes with quite
small $\G p$ cross sections, such as $\Upsilon$-production. For more
details see \cite{Felix97}.

Photo-induced processes are also of practical importance as they are
a serious source of beam loss as they lead in general to a change of the
charge-to-mass ratio of the nuclei. Especially the cross section for
the excitation of the giant dipole resonance, a collective mode of the
nucleus, is rather large for the heavy systems (of the order of 100b).
The cross section scales approximately with $Z^{10/3}$. The
contribution nucleon resonances (especially the $\Delta$ resonance)
has also been confirmed experimentally in fixed target experiments
with 60 and~200 GeV/A (heavy ions at CERN, ``electromagnetic
spallation'') \cite{BrechtmannH88a,BrechtmannH88b,PriceGW88}. For
details of these aspects, we refer the reader to
\cite{KraussGS97,VidovicGS93,BaltzRW96,BaurB89}, where scaling laws,
as well as detailed calculations for individual cases are given.

\subsection{Photon-Photon Physics at various invariant mass scales}
\label{ssec_ggphysics}

Up to now photon-photon scattering has been mainly studied at $\EPEM$
colliders. Many reviews \cite{BudnevGM75,KolanoskiZ88,BergerW87}
as well as conference reports
\cite{Amiens80,SanDiego92,Sheffield95,Egmond97,Freiburg99} exist. The
traditional range of invariant masses has been the region of mesons,
ranging from $\pi^0$ ($m_{\pi^0}=135$~MeV) up to about $\eta_c$
($m_{\eta_c}=2980$~MeV). Recently the total $\GG\rightarrow$~hadron
cross-section has been studied at LEP2 up to an invariant mass range
of about 70~GeV \cite{L3:97}. We are concerned here mainly with the
invariant mass region relevant for LHC (see the
$\GG$-luminosity figures below). Apart from the production of $\EPEM$
(and $\MUPM$) pairs, the photons can always be considered as
quasireal. The cross section for
virtual photons deviates from the one for real photons only for $Q^2$,
which are much larger then the coherence limit $Q^2\lesssim 1/R^2$
(see also the discussion in \cite{BudnevGM75}). For real photons
general symmetry requirements restrict the possible final states, as
is well known from the Landau-Yang theorem. Especially
it is impossible to produce spin 1 final states. In $\EPEM$
annihilation only states with $J^{PC}=1^{--}$ can be produced
directly. Two photon collisions give access to most of the $C=+1$
mesons.

In principle $C=-1$ vector mesons can be produced by the fusion of
three (or, less important, five, seven, \dots) equivalent
photons. This cross section scales with $Z^6$. But it is smaller than
the contribution coming from $\G$-A collisions, as discussed above,
even for nuclei with large $Z$ (see \cite{BaurHT98}).

The cross section for $\GG$-production in a heavy ion collision
factorizes into a $\GG$-luminosity function and a cross-section
$\sigma_{\GG}(W_{\GG})$ for the reaction of the (quasi)real photons
$\GG \rightarrow f$, where $f$ is any final state of interest (see
Eq.~(\ref{eq_sigmaAA}).  When the final state is a narrow resonance,
the cross-section for its production in two-photon collisions is given
by
\BE
\sigma_{\GG\rightarrow R}(M^2) =
 8 \pi^2 (2 J_R+1) \Gamma_{\GG}(R) \delta(M^2-M_R^2)/M_R ,
\label{eq_nres}
\EE
where $J_R$, $M_R$ and $\Gamma_{\GG}(R)$ are the spin, mass and
two-photon width of the resonance $R$. This makes it easy to calculate
the production cross-section $\sigma_{AA\rightarrow AA+R}$ of a
particle in terms of its basic properties.

\begin{figure}[tbhp]
\begin{center}
\resizebox{8cm}{!}{\includegraphics{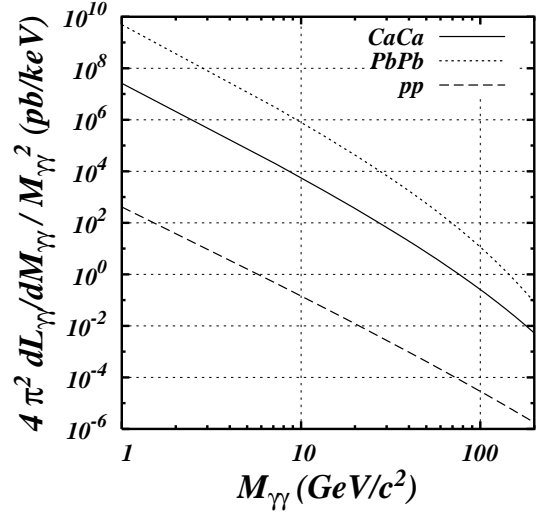}}
\end{center}
\caption{\it
The universal function $4\pi^2 dL_{\GG}/dM_{\GG} /M_{\GG}^2$ is
plotted for different ion species at LHC. We use $R=1.2 A^{1/3}$~fm and
$\gamma=2950$, 3750 and 7000 for Pb-Pb, Ca-Ca and $p$-$p$, respectively.
}
\label{fig_sigmagamma}
\end{figure}

In Fig.~\ref{fig_sigmagamma} the function $4\pi^2 dL_{\GG}/dM /M^2$,
which is universal for a produced resonances, is plotted for various
systems. It can be directly used to calculate the cross-section for
the production of a resonance 
$R$ with the formula 
\BE
\sigma_{AA\rightarrow AA+R}(M) = (2 J_R +1) \Gamma_{\GG} \frac{4 \pi^2
dL_{\GG}/dM}{M^2} .
\label{eq_aar}
\EE
%
%

We will now give a general
discussion of possible photon-photon physics at relativistic heavy ion
colliders. Invariant masses up to several GeV can be reached at RHIC
and up to about 100 GeV at LHC.

We can divide our discussion into the following two main subsections:
Basic QCD phenomena in $\GG$-collisions (covering the range of meson,
meson-pair production, etc.) and $\GG$-collisions as a tool for new
physics, especially at very high invariant masses.
An interesting topic in itself is the $e^+$-$e^-$ pair 
production. The fields are strong enough to produce multiple pairs in a
single collisions. A discussion of this subject together with calculations 
within the semiclassical approximation can be found in 
\cite{Baur90,HenckenTB95a,HenckenTB95b,AlscherHT97}

\subsection{Basic QCD phenomena in $\GG$-collisions}
\label{ssec_basicqcd}

\subsubsection{Hadron spectroscopy: Light and heavy quark spectroscopy}

One may say that photon-photon collisions provide an independent view
of the meson and baryon spectroscopy. They provide powerful
information on both the flavor and spin/angular momentum internal
structure of the mesons. Much has already been done at
$\EPEM$ colliders. Light quark spectroscopy is very
well possible at RHIC, benefiting from the high
$\GG$-luminosities. Detailed feasibility studies exist
\cite{KleinS97a,KleinS97b,KleinS95a,KleinS95b}.  In these studies, $\GG$
signals and backgrounds from grazing nuclear and beam gas collisions
were simulated with both the FRITIOF and VENUS Monte Carlo codes. The
narrow $p_\perp$-spectra of the $\GG$-signals provide a good
discrimination against the background, see also the discussion of a
possible trigger in \ref{ssec_selecting} below. The possibilities to
produce these mesons at the LHC have been discussed in detail in the 
FELIX LoI \cite{Felix97}. Also discussed there are how to isolate
$\GG$ events in ion-ion collisions and applications to basic QCD
phenomena like $C=+1$ meson production, vector meson pair production
and total hadronic $\GG$ cross sections. Rates are given and possible
triggers are discussed. In addition photon-Pomeron and
Pomeron-Pomeron processes are discussed. The general conclusion is
that all these processes are very promising tools for vector meson
spectroscopy.

In particular the absence of meson production via $\GG$-fusion is also
of great interest for glueball search. The two-photon width of a
resonance is a probe of the charge of its constituents, so the
magnitude of the two-photon coupling can serve to distinguish quark
dominated resonances from glue-dominated resonances (``glueballs'').
In $\GG$-collisions, a glueball can only be produced via the
annihilation of a $q\bar q$ pair into a pair of gluons, whereas a
normal $q\bar q$-meson can be produced directly. Therefore we expect
the ratio for the production of a glueball $G$ compared to a normal
$q\bar q$ meson $M$ to be
\BE
\frac{\sigma(\GG \rightarrow M)}{\sigma(\GG \rightarrow G)}
=
\frac{\Gamma(M \rightarrow \GG)}{\Gamma(G \rightarrow \GG)}
\sim
\frac{1}{\alpha_s^2} ,
\EE
where $\alpha_s$ is the strong interaction coupling constant. On the
other hand glueballs are most easily produced in a glue-rich
environment, for example, in radiative $J/\Psi$ decays, $J/\Psi
\rightarrow \gamma gg$. In this process we expect the ratio of the
cross section to be
\BE
\frac{\Gamma(J/\Psi \rightarrow \G G)}{\Gamma(J/\Psi \rightarrow \G
M)} \sim \frac{1}{\alpha_s^2} .
\EE
A useful quantity to describe the gluonic character of a mesonic state
X is therefore the so called ``stickiness'' \cite{Cartwright98}, defined as
\BE
S_X = \frac{\Gamma(J/\Psi \rightarrow \G X)}{\Gamma(X \rightarrow
\G \G)} .
\EE
One expects the stickiness of all mesons to be comparable, while for
glueballs it should be enhanced by a factor of about $ 1/\alpha_s^4 \sim 20$.
In a recent reference \cite{Godang97} results of the search for $f_J
(2220)$ production in two-photon interactions were presented. There a
very small upper limit for the product of $\Gamma_{\GG} B_{K_sK_s}$
was given, where $B_{K_s K_s}$ denotes the branching fraction of
its decay into $K_s K_s$.  From this it was concluded that this is a
strong evidence that the $f_J(2220)$ is a glueball.

For charmonium production, the two-photon width $\Gamma_{\GG}$ of
$\eta_c$ (2960 MeV, $J^{PC} = 0^{-+}$) is known from experiment
\cite{PDG98}. But the two-photon widths of $P$-wave charmonium states
have been measured with only modest accuracy. Two photon widths of
$P$-wave charmonium states can be estimated following the PQCD
approach \cite{Bodwin92}. Similar predictions of the bottonia two
photons widths can be found in \cite{Kwong88,Consoli94}. For RHIC the
study of $\eta_c$ is a real challenge \cite{KleinS97b}; the
luminosities are falling and the branching ratios to experimentally
interesting channels are small. 

In Table~\ref{tab_ggmeson} (adapted from table~2.6 of \cite{Felix97})
the two-photon production cross-sections
for $c\bar c$ and $b \bar b$ mesons in the rapidity range $|Y|<7$ are
given. Also given are the number of events in a $10^7$ sec run with
the ion luminosities of $4\times 10^{30}$cm${}^{-2}$s${}^{-1}$ for
Ca-Ca and $10^{26}$cm${}^{-2}$s${}^{-1}$ for Pb-Pb. Millions of
$C$-even charmonium states will be produced in coherent two-photon
processes during a standard $10^7$~sec heavy ion run at the LHC. The
detection efficiency of charmonium events has been estimated as 5\%
for the forward-backward FELIX geometry \cite{Felix97}, i.e., one can
expect detection of about $5\times 10^3$ charmonium events in Pb-Pb
and about $10^6$ events in Ca-Ca collisions. This is two to three
orders of magnitude higher than what is expected during five years of
LEP200 operation. Experiments with a well-equipped central detector
like CMS on the other hand should provide a much better
efficiency. Further details, also on experimental cuts, backgrounds and
the possibilities for the study of $C$-even bottonium states are given
in \cite{Felix97}. 
%
%
\begin{table}[hbt]
\begin{center}
\begin{tabular}{|l|c|c|r|c|c|c|c|}
\hline
State   & Mass,     & $\Gamma_{\GG}$ &
              \multicolumn{2}{|c}{$\sigma (AA\to AA+X)$} &
              \multicolumn{2}{|c|}{rates per $10^7$ sec}\\
\cline{4-5}
  & MeV     &  keV  &      Pb-Pb  &                Ca-Ca  
  & Pb-Pb & Ca-Ca \\
\hline
~~~$\pi_0$  & 134 & $8\times10^{-3}$   & 46 mbarn& 210 $\mu$barn 
   & $4.6 \times 10^7$ & $8.4 \times 10^9$ \\
~~~$\eta$   & 547  & 0.46 & 20 mbarn& 100 $\mu$barn 
   & $2 \times 10^7  $ & $4.0 \times 10^9$ \\
~~~$\eta'$  & 958 & 4.2 & 25 mbarn  & 130 $\mu$barn 
   & $2.5 \times 10^7$ & $5.2 \times 10^9$ \\
~~~$f_2(1270)$   & 1275   & 2.4 & 25 mbarn & 133 $\mu$barn 
   & $2.5 \times 10^7$ & $5.2 \times 10^9$ \\
~~~$a_2(1320)$   & 1318   & 1.0 & 9.2 mbarn& 49 $\mu$barn 
   & $9.2 \times 10^6$ & $2.0 \times 10^9$ \\
~~~$f_2'(1525)$  & 1525   & 0.1 & 540 $\mu$barn& 2.9 $\mu$barn 
   & $5.4 \times 10^5$ & $1.2 \times 10^8$ \\
~~~$\eta_c$ & 2981 & 7.5 & 360 $\mu$barn & 2.1 $\mu$barn 
   & $3.6 \times 10^5$ & $8.4 \times 10^7$ \\
~~~$\chi_{0c}$& 3415& 3.3& 180 $\mu$barn &1.0 $\mu$barn 
   & $1.8 \times 10^5$ & $4.0 \times 10^7$ \\
~~~$\chi_{2c}$& 3556 & 0.8 & 74 $\mu$barn & 0.44 $\mu$barn 
   & $7.4 \times 10^4$ & $1.8 \times 10^7$ \\
~~~$\eta_b$   & 9366 & 0.43 &450 nbarn & 3.1 nbarn   
   & 450             & $1.2 \times 10^4$  \\
~~~$\eta_{0b}$& 9860 & $2.5\times10^{-2}$ & 21 nbarn & 0.15 nbarn 
   & 21              & 6000 \\
~~~$\eta_{2b}$& 9913 & $6.7\times10^{-3}$ & 28 nbarn & 0.20 nbarn 
   & 28              & 8000 \\
\hline
\end{tabular}
\end{center}
\caption{\it
Mass, and $\GG$-widths used to calculate the cross section for meson 
production for Pb-Pb and Ca-Ca collisions at CMS.
Masses and widths are taken from \cite{PDG98} and \cite{Felix97}. The
beam luminosities used are $10^{26}$cm${}^{-2}$s${}^{-1}$ for Pb-Pb
and $4\times10^{30}$cm${}^{-2}$s${}^{-1}$ for Ca-Ca.}
\label{tab_ggmeson}
\end{table}

\subsubsection{Vector-meson pair production. Total hadronic
cross-section} 

There are various mechanisms to produce hadrons in photon-photon
collisions. Photons can interact as point particles which produce
quark-antiquark pairs (jets), which subsequently hadronize. Often a
quantum fluctuation transforms the photon into a vector meson
($\rho$,$\omega$,$\phi$, \dots) (VMD component) opening up all the
possibilities of hadronic interactions .  In hard scattering, the
structure of the photon can be resolved into quarks and
gluons. Leaving a spectator jet, the quarks and gluon contained in the
photon will take part in the interaction.  It is of great interest to
study the relative amounts of these components and their properties.
%
%
\begin{figure}[tbhp]
\begin{center}
\resizebox{7.5cm}{!}{\includegraphics{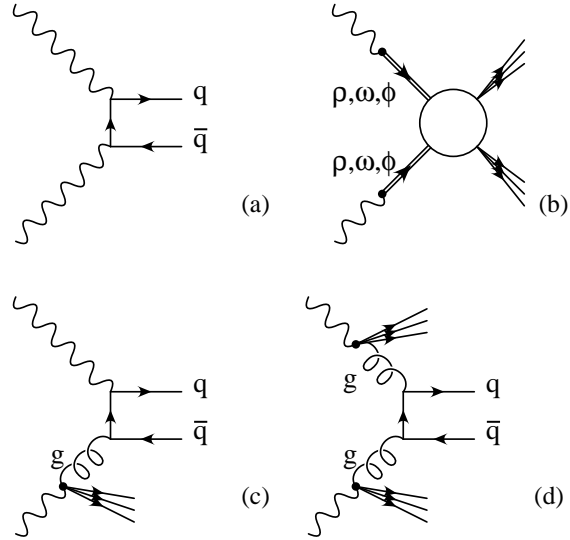}}
\end{center}

\caption{\it
Diagrams showing the contribution to the $\GG\rightarrow$hadron
reaction: direct mechanism (a), vector meson dominance (b), single (c)
and double (d) resolved photons.
}
\label{fig_resolved}
\end{figure}

The L3 collaboration recently made a measurement of the total hadron
cross-section for photon-photon collisions in the interval $5 GeV <
W_{\GG} < 75 GeV$ \cite{L3:97}. It was found that the $\GG
\rightarrow$hadrons cross-section is consistent with the universal
Regge behavior of total hadronic cross-sections.
The production of vector meson pairs can well be studied at RHIC with
high statistics in the GeV region \cite{KleinS97a}.  For the
possibilities at LHC, we refer the reader to \cite{Felix97} and 
\cite{BaurHTS98}, where also experimental details and simulations are
described.

\subsection{$\GG$-collisions as a tool for new physics}
\label{ssec_newphysics}

The high flux of photons at relativistic heavy ion colliders offers
possibilities for the search of new physics. This includes the
discovery of the Higgs-boson in the $\GG$-production channel or new
physics beyond the standard model, like supersymmetry or
compositeness.

Let us mention here the plans to build an $\EPEM$ linear collider.
Such future linear colliders will be used for $\EPEM$, $e\G$
and $\GG$-collisions (PLC, photon linear collider). 
The photons will be obtained by scattering of
laser photons (of eV energy) on high energy electrons ($\approx$ TeV
region) (see \cite{Telnov95}). Such photons in the TeV energy range
will be monochromatic and polarized. The physics program at such
future machines is discussed in \cite{ginzburg95}, it includes Higgs
boson and gauge boson physics and the discovery of new particles.

While the $\GG$ invariant masses which will be reached at RHIC will
mainly be useful to explore QCD at lower energies, the $\GG$ invariant
mass range at LHC --- up to about 100 GeV --- will open up new
possibilities.

A number of calculations have been made for a medium heavy standard
model Higgs \cite{DreesEZ89,MuellerS90,Papageorgiu95,Norbury90}. For
masses $m_H < 2 m_{W^\pm}$ the Higgs bosons decays dominantly into
$b\bar b$. Chances of finding the standard model
Higgs in this case are marginal \cite{BaurHTS98}.

An alternative scenario with a light Higgs boson was, e.g., given in
\cite{ChoudhuryK97} in the framework of the ``general two Higgs
doublet model''. Such a model allows for a very light particle in the
few GeV region. With a mass of 10~GeV, the $\GG$-width is about 0.1
keV. The authors of
\cite{ChoudhuryK97} proposed to look for such a light neutral Higgs
boson at the proposed low energy $\GG$-collider. We want to point out
that the LHC Ca-Ca heavy ion mode would also be very suitable for such
a search.

One can also speculate about new particles with strong coupling to the
$\GG$-channel. Large $\Gamma_{\GG}$-widths will directly lead to large
$\GG$ production cross-sections. We quote the
papers \cite{Renard83,BaurFF84}. Since the $\GG$-width of a resonance
is mainly proportional to the wave function at the origin, huge values
can be obtained for very tightly bound systems. Composite scalar
bosons at $W_{\GG}\approx 50$~GeV are expected to have $\GG$-widths of
several MeV \cite{Renard83,BaurFF84}. The search for such kind of
resonances in the $\GG$-production channel will be possible at
LHC.

In Refs. \cite{DreesGN94,OhnemusWZ94} $\GG$-processes at $pp$
colliders (LHC) are studied. It is observed there that non-strongly
interacting supersymmetric particles (sleptons, charginos,
neutralinos, and charged Higgs bosons) are difficult to detect in
hadronic collisions at the LHC. The Drell-Yan and gg-fusion mechanisms
yield low production rates for such particles. Therefore the
possibility of producing such particles in $\GG$ interactions at
hadron colliders is examined. Since photons can be emitted from
protons which do not break up in the radiation process, clean events
can be generated which should compensate for the small number. In
\cite{DreesGN94} it was pointed out that at the high luminosity of
$L=10^{34}$cm${}^{-2}$s${}^{-1}$ at the LHC($pp$), one expects about
16 minimum bias events per bunch crossing. Even the elastic $\GG$
events will therefore not be free of hadronic debris. Clean elastic
events will be detectable at luminosities below
$10^{33}$cm${}^{-2}$s${}^{-1}$. This danger of ``overlapping events''
has also to be checked for the heavy ion runs, but it will be much
reduced due to the lower luminosities.
Recent (unpublished) studies done for FELIX and ALICE show that the
chargino pair production can be detectable, if the lighest chargino
would have a mass below 60~GeV$/c^2$. Unfortunately recent chargino
mass limits set by LEP experiments already exclude the existence of
charginos on this mass range. Therefore the observation of
MSSM-particles in $\GG$-interactions in heavy ion collisions seems to
be hard to achieve.

Similar considerations for new physics were also made in connection
with the planned $eA$ collider at DESY (Hamburg). Again, the coherent
field of a nucleus gives rise to a $Z^2$ factor in the cross-section
for photon-photon processes in $eA$ collisions \cite{KrawczykL95}.

\subsection{Dilepton production}
\label{ssec_leptons}

Electrons (positrons) and to some extent also muons have a special
status, which is due to their small mass. They are therefore produced
more easily than other heavier particles and in the case of $\EPEM$
pair production also lead to new phenomena, like multiple pair
production. Due to their small mass and therefore large Compton wave
length (compared to the nuclear radius), the equivalent photon
approximation has to be modified when applied to them. For the
muon, with a Compton wavelength of about 2 fm, we expect the standard
equivalent photon approximation to be applicable, with only small
corrections. Both electrons and muons can be produced not only as free
particles but also into an atomic states bound to one of the ions, or
even as a bound state, positronium or muonium.

The special situation of the electron pairs can already be seen from
the formula for
the impact parameter dependent probability in lowest order. Using 
the equivalent photon approximation one obtains~\cite{BertulaniB88}
\BE
P^{(1)}(b) \approx \frac{14}{9 \pi^2} \left(Z \A\right)^4
\frac{1}{m_e^2 b^2} \ln^2 \left( \frac{\G_{ion} \delta}{2 m_e b}\right) ,
\label{eq_pbapprox}
\EE
where $\delta\approx 0.681$ and $\G_{ion}=2\G^2-1$ the Lorentz factor
in the target frame, one can see that at RHIC and LHC energies and for
impact parameters of the order of the Compton wave length $b\approx
1/m_e$, this probability exceeds one. Unitarity is restored by
considering the production of multiple pairs
\cite{Baur90,Baur90c,BestGS92,RhoadesBrownW91,HenckenTB95a}. To a good
approximation the multiple pair production can be described by a
Poisson distribution. The impact parameter dependent probability
needed in this Poisson distribution was calculated in lowest order in
\cite{HenckenTB95b,Guclu95}, the total cross section for the one-pair
production in \cite{Bottcher89}, for one and multiple pair production
in \cite{AlscherHT97}.  Of course
the total cross section is dominated by the single pair production as
the main contribution to the cross section comes from very large
impact parameters $b$.  On the other hand one can see that for impact
parameters $b$ of about $2R$ the number of electron-positron pairs
produced in each ion collision is about 5 (2) for LHC with $Z=82$
(RHIC with $Z=79$). This means that each photon-photon event ---
especially those at a high invariant mass --- which occur
predominantly at impact parameters close to $b \gtrsim 2 R$ --- is
accompanied by the production of several (low-energy) $\EPEM$ pairs.
%
%
\begin{figure}[tbhp]
\begin{center}
\resizebox{8cm}{!}{\includegraphics{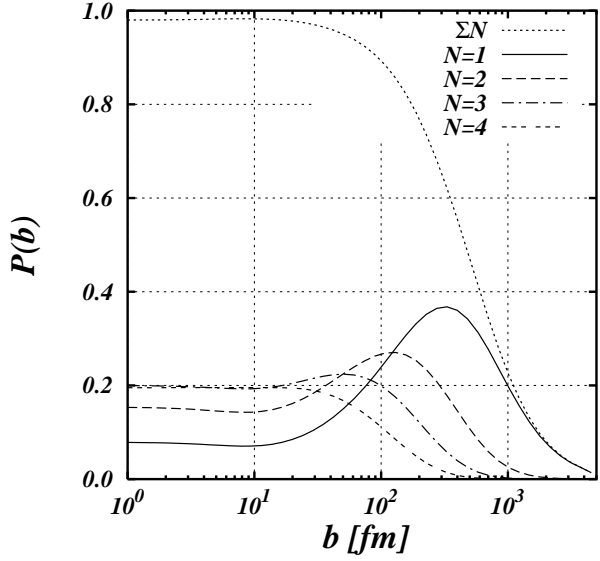}}
\end{center}
\caption{\it
The impact parameter dependent probability to produce $N$
$\EPEM$-pairs ($N=1,2,3,4$) in one collision is shown for the
 LHC ($\G=2950$,Pb-Pb). Also shown is the
total probability to produce at least one $\EPEM$-pair. One sees that at
small impact parameters multiple pair production dominates over
single pair production.
}
\label{fig_pbee1}
\end{figure}

\begin{figure}[tbh]
\begin{center}
\resizebox{7.5cm}{!}{\includegraphics{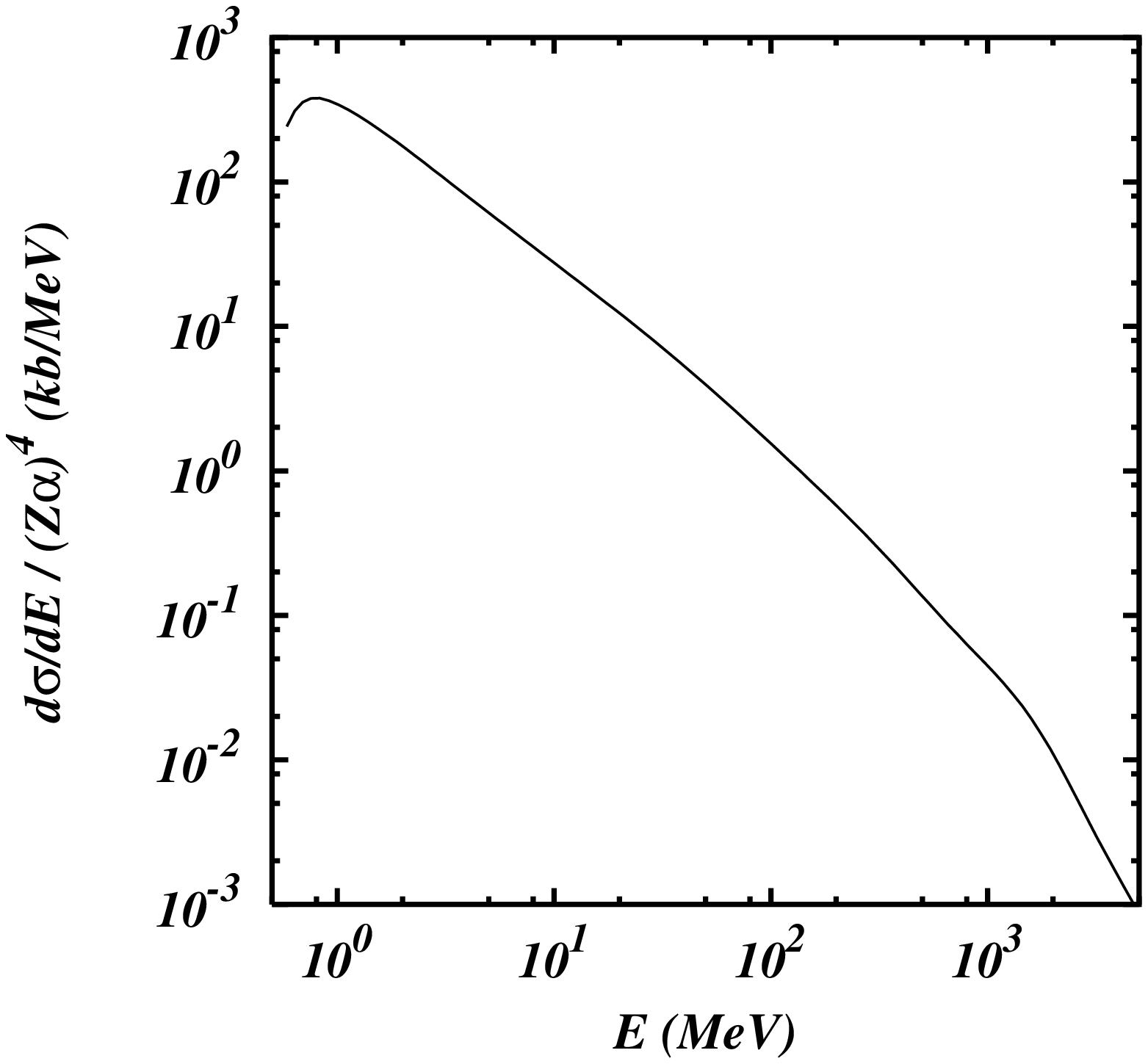}}
\resizebox{7.5cm}{!}{\includegraphics{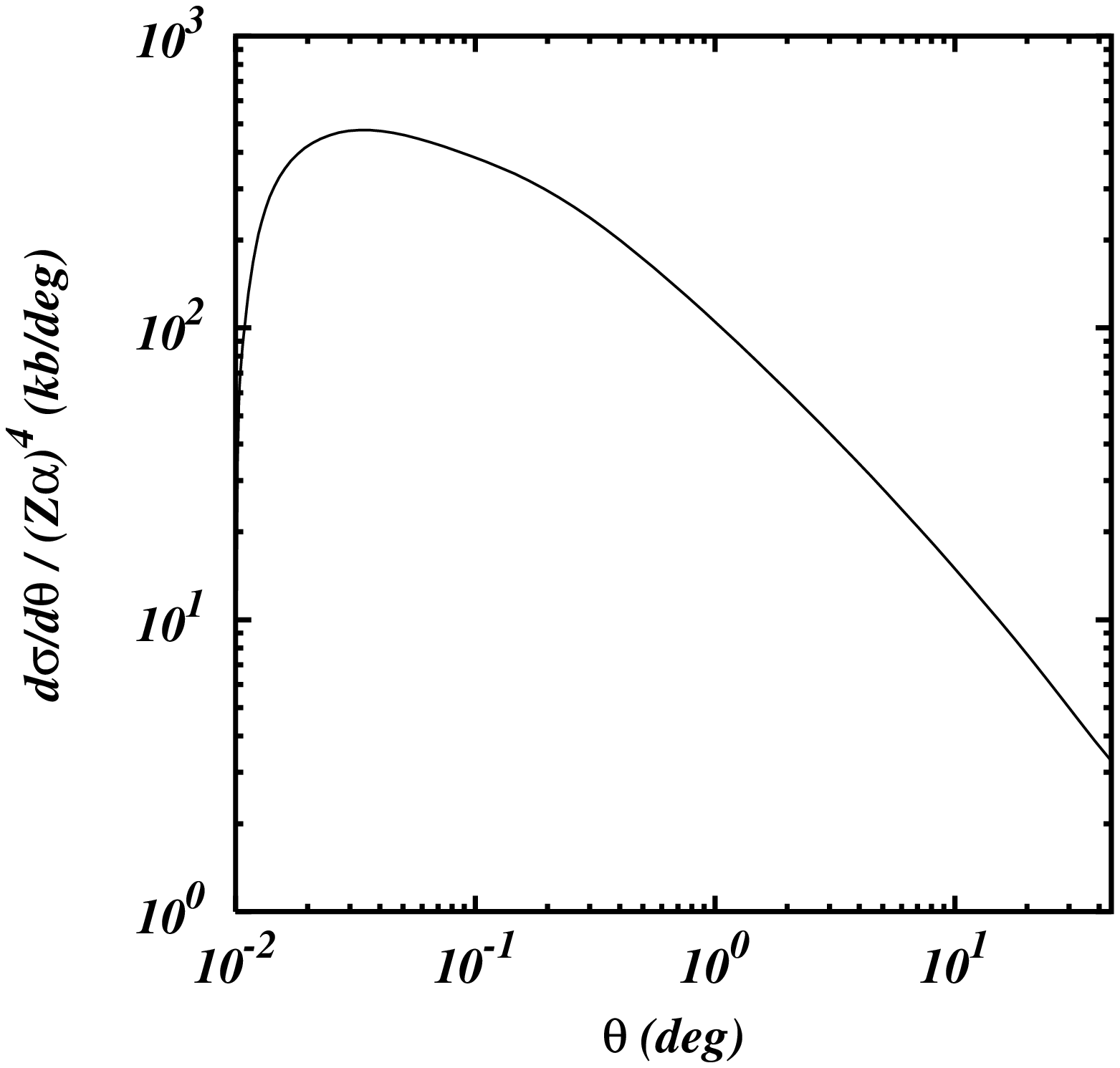}}
\end{center}
\caption{
Cross section for the $e^+e^-$ pair production as a function of the
energy (A) of either electron or positron and as a function of the
angle of the electron or positron with the beam axis (B). 
Most pairs are produced with energies between 2--5 MeV and in the very
forward or backward direction.
}
\label{fig_eee}
\end{figure}

As the total cross section for this process is huge (about 200~kbarn for
Pb-Pb at LHC), one has to take this process into account as a possible
background process. Most of the particles are produced at low
invariant masses (below 10 MeV) and into the very forward direction
(see Fig.~\ref{fig_eee}). High energetic electrons and positrons are
even more concentrated along the beam pipe, most of them therefore are
unobserved. On the other hand, a substantial amount of them is still
left at high energies, e.g., above 1~GeV. These QED pairs therefore
constitute a potential hazard for the detectors, see below in
Sec.~\ref{ssec_selecting}. One the other hand, they can also be useful
as a possible luminosity monitor, as discussed in \cite{Felix97,ShamovT98}.
%
%

Differential production probabilities for $\GG$-dileptons in central 
relativistic heavy ion collisions are calculated using the equivalent
photon approximation and an
impact parameter formulation and compared to Drell-Yan and thermal
ones in \cite{Baur92,BaurB93b,Baur92b}. The very low $p_\perp$ values
and the angular distribution of the pairs give a handle for their
discrimination.

Higher order corrections, e.g., Coulomb corrections, have to be taken 
into account for certain regions in the phase space.
A classical result for these higher-order
effects can be found in the Bethe-Heitler formula for the process
$Z+\gamma \rightarrow Z + e^+ + e^-$ 
\BE
\sigma = \frac{28}{9} Z^2 \A r_e^2 \left[ \ln \frac{2 \omega}{m_e} -
\frac{109}{42} - f(\ZA) \right] ,
\EE
with the higher-order term given by
\BE
f(\ZA) = (\ZA)^2 \sum_{n=1}^{\infty} \frac{1}{n(n^2+(\ZA)^2)}
\EE
and $r_e=\A/m_e$ is the classical electron radius. As far as total
cross sections are concerned the higher-order contributions tends to a
constant for $\omega \rightarrow \infty$. 
A systematic way to take leading terms of higher order
effects into account in $\EPEM$ pair production is pursued in
\cite{IvanovM97,Ivanov98} using Sudakov variables and the impact-factor
representation. They find a reduction of the single-pair
production cross section of the order of 10\%.
In contrast to this some papers have recently discussed
nonperturbative results using a light-cone approach
\cite{segevW97,EichmannRSW98,BaltzM98}. There it is found that the
single-pair production cross section is identical to the lowest order
result. A calculation of the change of multiple pair production
cross section due to such higher order effects 
can be found in \cite{henckenTB99}.

\subsubsection{Equivalent Muons}

\begin{figure}[tbhp]
\begin{center}
\resizebox{5cm}{!}{\includegraphics{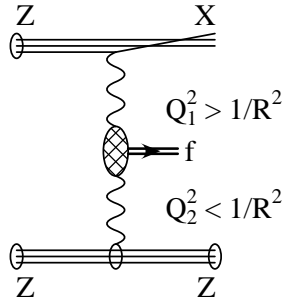}}
\end{center}
\caption{\it With $Q^2 < 1/R^2$ the photon is emitted coherently from
all ``partons'' inside the ion. For $Q^2 \gg 1/R^2$ the ``partonic''
structure of the ion is resolved.}
\label{fig_dis}
\end{figure}

Up to now only the production of dileptons was considered, for which
the four-momentum $Q^2$ of the photons was less than about $1/R^2$
(coherent interactions). There is another class of processes, where
one of the interactions is coherent ($Q^2 \le 1/R^2$) and the other
one involves a deep inelastic interaction ($Q^2\gg 1/R^2$), see
Fig.~\ref{fig_dis}. These processes are readily described using the
equivalent electron-- (or muon--, or tau--) approximation, as given,
e.g., in \cite{ChenZ75,BaierFK73}. The equivalent photon can be
considered as containing muons as partons, that is, consisting in part
of an equivalent muon beam. The equivalent muon number is given by 
\cite{ChenZ75}
\BE
f_{\mu/\G} ( \omega,x) = \frac{\alpha}{\pi}
\ln\left(\frac{\omega}{m_\mu}\right) \left( x^2 + (1-x)^2 \right),
\EE
where $m_\mu$ denotes the muon mass. The muon energy $E_\mu$ is given by
$E_\mu=x \omega$, where $\omega$ is the energy of the equivalent photon.
This spectrum has to be folded with the equivalent photon spectrum given by
\BE
f_{\G/Z}(u) = \frac{2\alpha}{\pi} \frac{Z^2}{u} \ln\left( \frac{1}{u
m_A R}\right)
\EE
for $u<u_{max}=\frac{1}{R m_A}$. 
The deep inelastic lepton-nucleon scattering can now be calculated in
terms of the structure functions $F_1$ and $F_2$ of the nucleon.
The inclusive cross section for the  deep-inelastic scattering of the
equivalent muons is therefore given by
\BE
\frac{d^2\sigma}{dE' d\Omega} = \int dx_1 f_{\mu/Z}(x_1)
\frac{d^2\sigma}{dE' d\Omega} (x_1)
\EE
where $\frac{d^2\sigma}{dE' d\Omega} (x_1)$
can be calculated from the usual invariant variables in 
deep inelastic lepton scattering (see, e.g., Eq. 35.2 of \cite{PDG96})
The lepton is scattered to an angle $\theta$ with an energy $E'$. The
equivalent muon spectrum of the heavy ion is obtained as
\BE
f_{\mu/Z} (x_1) = \int_{x_1}^{u_{max}} du f_{\G/Z} (u) f_{\mu/\G}(x_1/u).
\EE
This expression can be calculated
analytically and work on this is in progress \cite{BaurHT99}.
%
%
%

Such events are characterized by a single muon with an energy $E'$ and
scattering angle $\theta$. The accompanying muon of opposite charge,
as well as the remnants of the struck nucleus, will scatter to small
angles and remain unobserved. The hadrons scattered to large angles
can be observed, with total energy $E_h$ and momentum in the beam
direction of $p_{zh}$. Using the Jacquet-Blondel variable $y_{JB}$ the
energy of the equivalent muon can in principle be reconstructed as
\BE
E_\mu = \frac{1}{2} \left( E_h - p_{zh} + E' (1-\cos\theta)\right)
\EE

This is quite similar to the situation at HERA, with the difference
that the energy of the lepton beam is continuous, and its energy has
to be reconstructed from the kinematics (How well this can be done in
practice remains to be seen).

\subsubsection{Radiation from $\EPEM$ pairs}

\begin{figure}[tbhp]
\begin{center}
\resizebox{8cm}{!}{\includegraphics{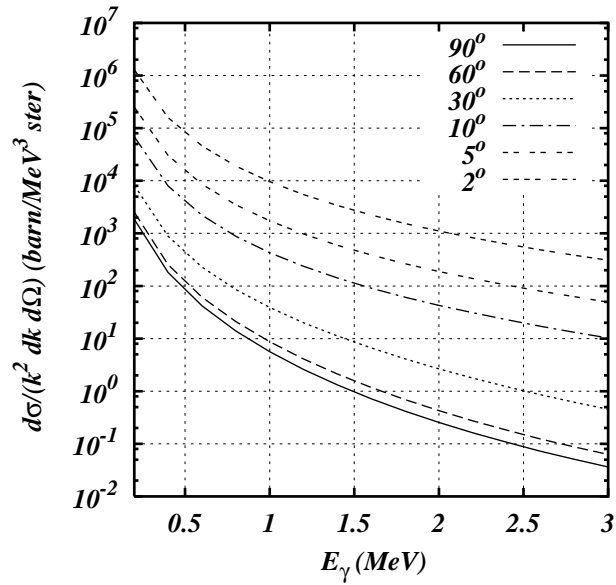}}
\end{center}
\caption{\it The energy-dependence of bremsstrahlungs-photons from
$e^+ e^-$ pair production is shown for different angles. We show
results for Pb-Pb collisions at LHC.} 
\label{fig_brems}
\end{figure}

The bremsstrahlung in peripheral relativistic heavy ion collisions was
found to be small, both for real \cite{BertulaniB88} and virtual
\cite{MeierHTB98} bremsstrahlung photons. This is due to the large
mass of the heavy ions.
Since the cross section for $\EPEM$ pair production is so large, one
can expect to see sizeable effects from the radiation of these light
mass particles. In the soft photon limit (see, e.g., \cite{Weinberg97}) one
can calculate the cross section for soft
photon emission of the process as
\BE
Z + Z \rightarrow Z + Z + e^+ + e^- + \gamma
\EE
as
\BE
d\sigma(k,p_-,p_+) = 
- e^2 \left[ \frac{p_-}{p_- k} - \frac{p_+}{p_+ k} \right]^2
\frac{d^3k}{4 \pi^2 \omega} d\sigma_0(p_+,p_-)
\EE
where $d\sigma_0$ denotes the cross section for the $e^+ e^-$ pair
production in heavy ion collisions. An alternative approach is done
by using the equivalent photon approximation (EPA) and
calculating the exact lowest order matrix element for the process
$$
\gamma + \gamma \rightarrow e^+ + e^- + \gamma.
$$
In Fig.~\ref{fig_brems} we show results of calculations for low energy 
photons. For this we have used the exact lowest order QED process in the
equivalent photon approximation \cite{HenckenTB99b}.
%
%

These low energy photons might constitute a background for the
detectors. Unlike the low energy electrons and positrons, they are of
course not bent away by the magnets. The angular distribution of the
photons also peak at small angles, but again a substantial amount is
still left at larger angles, even at $90^o$. The typical energy of
these low energy photons is of the order of several MeV, i. e., much
smaller than the expected level of the energy equivalent noise in the
CMS ECALs \cite{CMSTechProp}.

\subsubsection{Bound-free Pair Production}

The bound-free pair production, also known as electron-pair production
with capture, is a process, which is also of practical importance in
the collider. It is the process, where a pair is produced but with the
electron not as a free particle, but into an atomic bound state of one
of the nuclei. As this changes the charge state of the nucleus, it is
lost from the beam. Together with the electromagnetic dissociation of
the nuclei (see Sec.~\ref{ssec_ga}) these two processes are the
dominant loss processes for heavy ion colliders.

In \cite{BertulaniB88} an approximate value for this cross section is
given as
\BE
\sigma_{capt}^K \approx \frac{33\pi}{10} Z_1^2 Z_2^6 \A^6 r_e^2
\frac{1}{\exp(2\pi Z_2 \A) -1} \left[ \ln\left(\G_{ion} \delta / 2\right) -
\frac{5}{3}\right] ,
\label{eq_capture}
\EE
where $\G_{ion}=2 \G^2 - 1$ is the Lorentz factor of the ion
in the rest frame of the other ion and only capture to the $K$-shell
is included. The cross section for all higher shells is expected to be
of the order of 20\% of this cross section (see Eqs 7.6.23 and 24 of
\cite{BertulaniB88}).

The cross section in Eq.~(\ref{eq_capture}) is of the form 
\BE
\sigma= C \ln \G_{ion} + D.
\label{eq_lnAB}
\EE
This form has been found to be a universal one at
sufficient high values of $\G$. The constant $C$ and $D$ then only
depend on the type of the target.

The above cross section was found making use of the equivalent photon
approximation (EPA) and also using
approximate wave function for bound state and continuum. More precise
calculations exist 
\cite{BaltzRW92,BaltzRW93,BeckerGS87,AsteHT94,AggerS97,RhoadesBrownBS89}
in the literature. Recent calculations within DWBA for high values of
$\G$ have shown that the exact first order results do not differ
significantly from EPA results \cite{MeierHHT98,BertulaniB98}.
Parameterizations for $C$ and $D$\cite{BaltzRW93,AsteHT94} for typical
cases are given in Table~\ref{tab_capture}.
%
%
\begin{table}[tbhp]
\begin{center}
\begin{tabular}{|c|c|c|c|c|}
\hline
Ion & $C$ & $D$ & $\sigma(\G=106)$ & $\sigma(\G=2950)$ \\
\hline
Pb & $15.4$barn & $-39.0$barn       & 115 barn     & 222 barn \\
Au & $12.1$barn & $-30.7$barn       &  90 barn     & 173 barn \\
Ca & $1.95$mbarn & $-5.19$mbarn     & 14 mbarn     & 27.8 mbarn\\
O  & $4.50\mu$barn & $-12.0\mu$barn & 32 $\mu$barn & 64.3 $\mu$barn \\
\hline
\end{tabular}
\end{center}
\caption{\it
Parameters $C$ and $D$ (see Eq.~(\ref{eq_lnAB})) 
as well as total cross sections for the bound-free pair production for
RHIC and LHC. The parameters are taken from 
\protect\cite{AsteHT94}.}
\label{tab_capture}
\end{table}

For a long time the effect of higher order and nonperturbative
processes have been under investigation. At lower beam energies, in the
region of few GeV per nucleon, coupled channel calculations have
indicated for a long time, that these give large contributions,
especially at small impact parameters. Newer calculation tend to
predict considerably smaller values, of the order of the first order
result and in a recent article Baltz \cite{Baltz97} finds in the limit
$\G\rightarrow \infty$ that contributions from higher orders are even
slightly smaller than the first order results.

The bound-free pair production was measured in two recent experiments
at the SPS, at fixed target $\G=168$ \cite{Krause98} and at fixed target
$\G \approx 2$ \cite{Belkacem93,Belkacem94}. Both experiments found good 
agreement between measurement and calculations.

We note that electron and positron can also form a bound state,
positronium. This is in analogy to the $\GG$-production of mesons
($q\bar q$ states) discussed in Sec.~\ref{ssec_ggphysics}. 
With the known width of the parapositronium $\Gamma((\EPEM)_{n=1}
{}^1S_0 \rightarrow \GG) = m c^2 \A^5 /2$, the photon-photon
production of this bound state was calculated in \cite{Baur90b}.  The
production of orthopositronium, $n=1 {}^3S_1$ was calculated recently
\cite{Ginzburg97}. 

As discussed in Sec.~\ref{ssec_ggphysics} the production of
orthopositronium is only suppressed by the factor $(\ZA)^2$, which is
not very small. Therefore one expects that both kind of positronium
are produced in similar numbers. Detailed calculation show that the
three-photon process is indeed not much smaller than the two-photon
process \cite{Ginzburg97,Gevorkyan98}.

\subsection{Event rates at CMS}
\label{ssec_evrate}
An overview of the expected event rate for a number of different
photon-photon reactions to either discrete states or continuum states
is given in the following figures. The right hand axes shows both the
number of events per second and per one-year run time (assuming $10^7$
sec per year). We use beam luminosities of
$10^{26}$cm${}^{-2}$s${}^{-1}$ for Pb-Pb and
$4\times10^{30}$cm${}^{-2}$s${}^{-1}$ for Ca-Ca.
The resonances have been calculated using the masses and photon-decay
widths as given in table~\ref{tab_ggmeson}. For the calculation of
the rate for a standard model Higgs boson $H_{SM}$, we use the approach as
discussed in \cite{DreesEZ89}. $H'$ denotes a nonstandard Higgs as
given in the ``general two-Higgs doublet model'' in
\cite{ChoudhuryK97}. As its photon-photon decay width is rather weakly 
dependent on its mass in the relevant mass region, we have used a constant 
value of 0.1 keV in our calculations.

The total hadronic cross section $\sigma_{\GG}($hadron$)$ was used in
the form \cite{L3:97}:
\begin{equation}
\sigma_{\GG}(\mbox{hadron}) = A (s/s_0)^\epsilon + B (s/s_0)^{-\eta}
\end{equation}
with $s_0=1$GeV${}^2$, $\epsilon=0.079$, $\eta=0.4678$, $A=173$ nbarn,
$B=519$ nbarn. For the dilepton and $q\bar q$ production via $\GG$, 
we have used the lowest order QED expression for pointlike fermions.
For the quark masses we use $m_c=1.3$~GeV and $m_b=4.6$~GeV
\cite{DreesKZZ93} (In this reference QCD corrections are also given). 
Of course these cross section only correspond to the ``direct
mechanism'' (see Figure~\ref{fig_resolved} above). In addition there
will be also events coming from resolved processes as well as vector
meson dominace \cite{SchulerS95,SchulerS96,SchulerS97}. This explains 
the much larger total hadronic cross section compared to the cross
section dye to the ``direct mechanism''. These two-quark processes
will be visible as two-jet events.
%
%
\begin{figure}[tbhp]
\begin{center}
\resizebox{10cm}{!}{\includegraphics{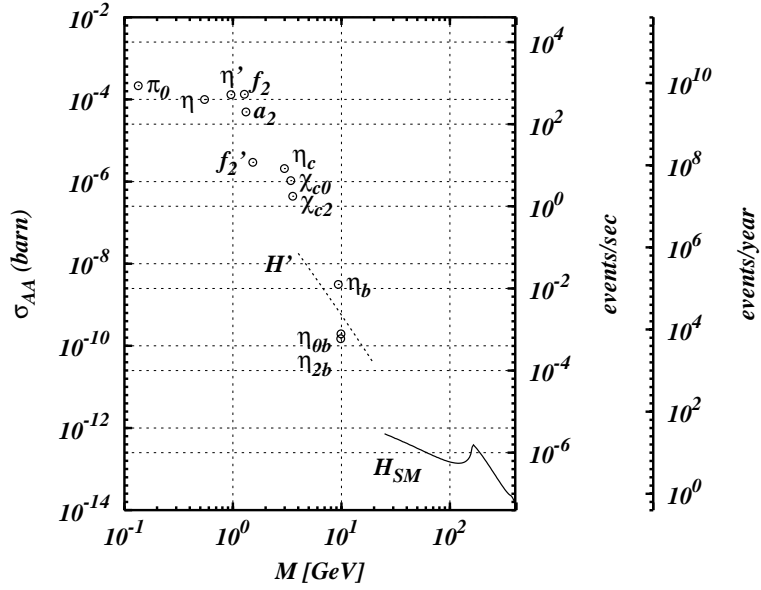}}
\end{center}
\caption{\it Overview of the total cross section and production rates 
(both per second and per one year run, assuming 1 year = $10^7$ sec) 
of different resonances in Ca-Ca collisions at the CMS. We have used
the parameters as given in the text and in
table~\protect\ref{tab_ggmeson}.}
\label{fig_caexres}
\end{figure}
%
%
\begin{figure}[tbhp]
\begin{center}
\resizebox{10cm}{!}{\includegraphics{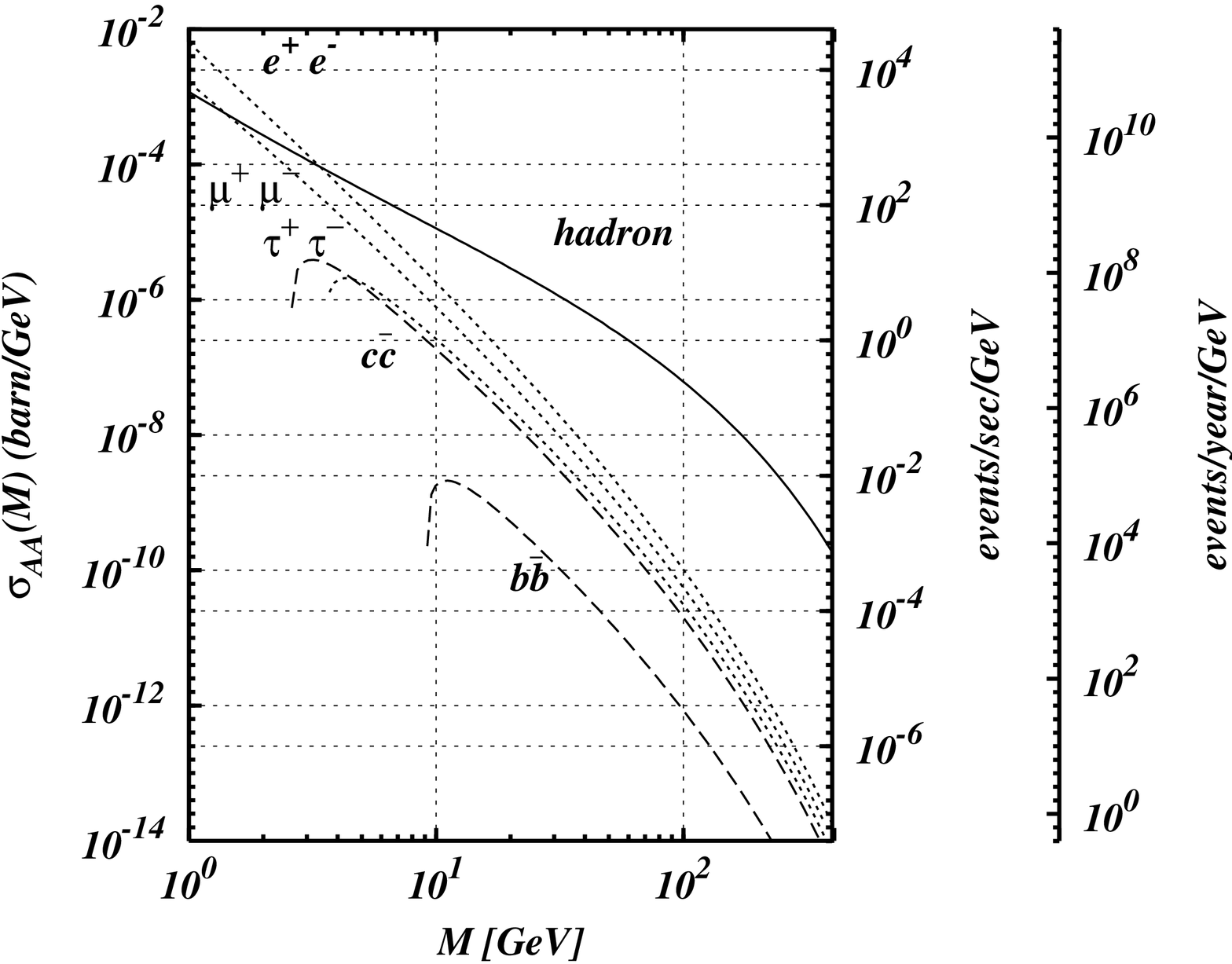}}
\end{center}
\caption{\it Overview of the total cross section and production rates 
(both per second and per one year run, assuming 1 year = $10^7$ sec) 
per GeV for different dilepton and
$q\bar q$ production for Ca-Ca collisions at CMS. Also shown is the total 
hadronic cross section. The parameters used are given in the text.}
\label{fig_caexcnt}
\end{figure}
%
%
\begin{figure}[tbhp]
\begin{center}
\resizebox{10cm}{!}{\includegraphics{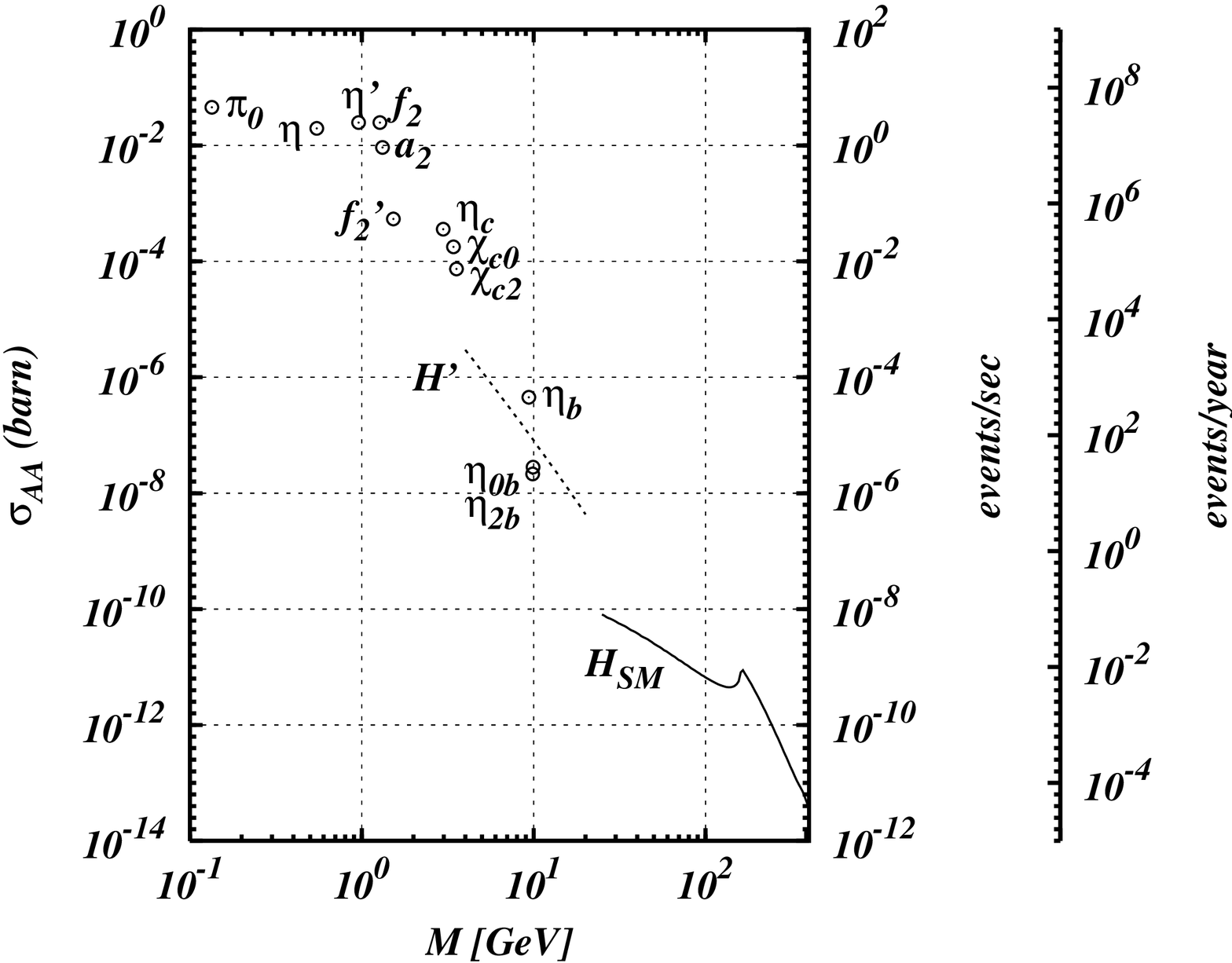}}
\end{center}
\caption{\it Overview of the total cross section and production rates 
(both per second and per one year run, assuming 1 year = $10^7$ sec)
 of different resonances in Pb-Pb collisions at the CMS. We have used
the parameters as given in the text and in
table~\protect\ref{tab_ggmeson}.} 
\label{fig_pbexres}
\end{figure}
%
%
\begin{figure}[tbhp]
\begin{center}
\resizebox{10cm}{!}{\includegraphics{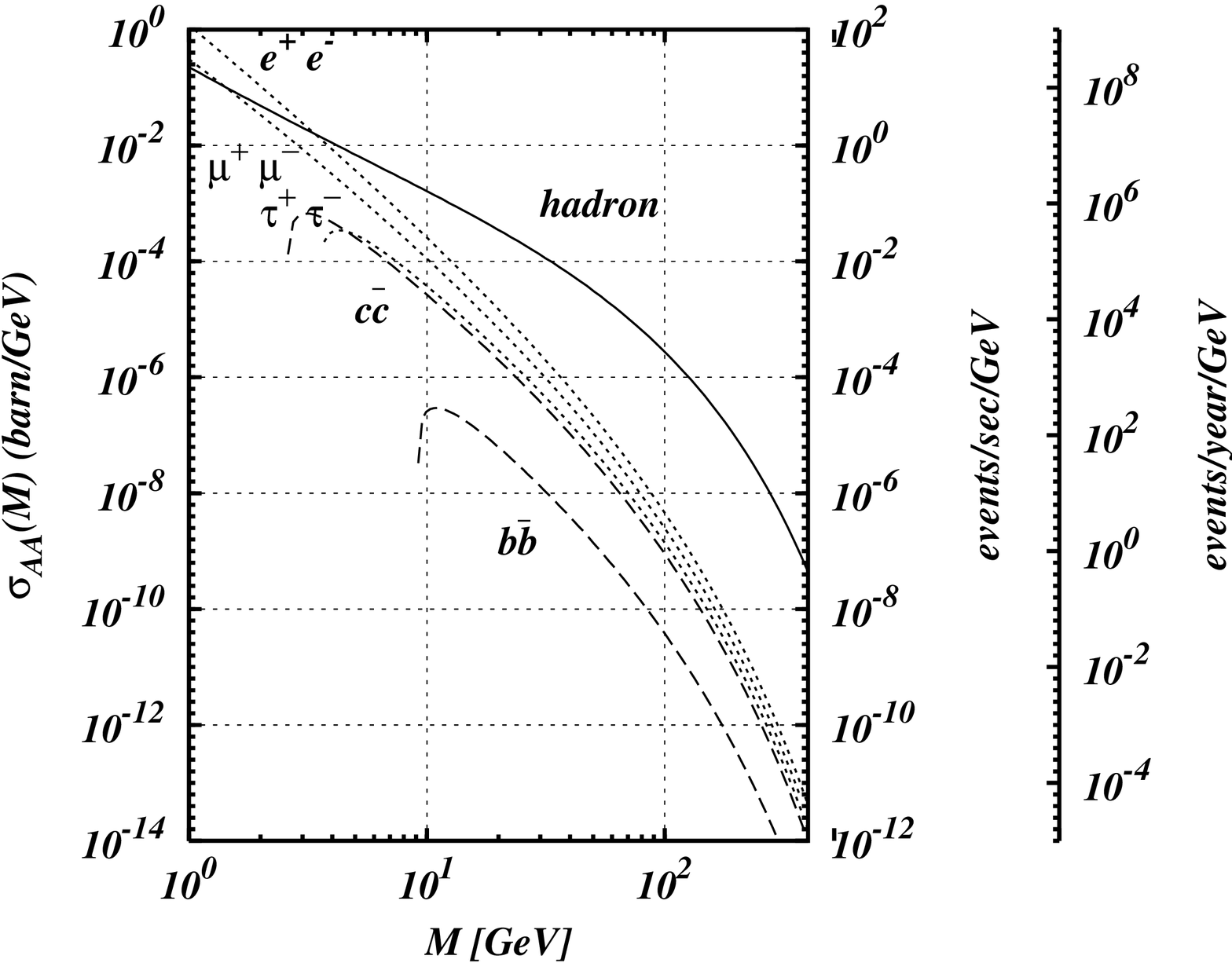}}
\end{center}
\caption{\it Overview of the total cross section and production rates 
(both per second and per one year run, assuming 1 year = $10^7$ sec) 
per GeV for different dilepton and
$q\bar q$ production for Pb-Pb collisions at CMS. Also shown is the total 
hadronic cross section. The parameters used are given in the text.}
\label{fig_pbexcnt}
\end{figure}

\subsection{Selecting $\GG$-events}
\label{ssec_selecting}

The $\GG$-luminosities are rather large, but the $\GG \rightarrow X$
cross sections are small compared to their hadronic counterparts,
therefore, e.g., the total hadronic production cross section for all
events is still dominated by hadronic events. This makes it necessary
to have an efficient trigger to distinguish photon-photon events from
hadronic ones.

There are some characteristic features that make such a trigger
possible. $\GG$-events are characterized by the fact that both nuclei
remain intact after the interaction. Therefore a $\GG$-event will have
the characteristic of a low multiplicity in the central region and no
event in the very forward or backward direction (corresponding to
fragments of the ions). The momentum transfer and energy loss for each
ion are too small for the ion to leave the beam.
It should be noted that in a $\GG$ interaction with an invariant mass of
several GeV leading to hadronic final states, quite a few particles
will be produced, see , e.g., \cite{L3:97}

A second characteristic is the small transverse momenta of the
produced system due to the coherence condition $q_\perp < 1/R\approx
50$~MeV. If one is able to make a
complete reconstruction of the momenta of all produced particles with
sufficient accuracy, this
can be used as a very good suppression against grazing collisions. As
the strong interaction is short ranged, it has normally a much broader
distribution in the transverse momenta. A calculation using the PHOJET
event generator \cite{EngelRR97} to study processes in central and
grazing collisions by Pomeron-exchange found an average transverse
momentum of about 450 MeV, about a factor of 10 larger than the
$\GG$-events. In a study for the STAR experiment \cite{NystrandK97} it was
also found that triggering for small transverse momenta is an
efficient method to reduce the background coming from grazing
collisions.

Another question that has to be addressed is the importance of
diffractive events, that is, Photon-Pomeron and Pomeron-Pomeron
processes in ion collisions.  From experiments at HERA one knows that
the proton has a large probability to survive intact after these
collisions. The theoretical situation unfortunately is not very clear
for these high energies and especially for nuclei as compared to
nucleons. Some calculations within the dual parton model have been
made and were interpreted as an indication that Photon-Pomeron and
Pomeron-Pomeron events are of the same size or even larger than
photon-photon events \cite{EngelRR97}. But these calculations were done
without requiring the condition to have intact nuclei in the final
state.  As the nuclei are bound only rather weakly and as mentioned
above the average momentum transfer to the nucleus is of the order of
200 MeV, it is very likely that the nucleus will break up in such a
collision. First estimates based on this model indicate, that this
leads to a substantial suppression of diffractive events, favoring
again the photon-photon events.

The cross section ratio of photon-photon to Pomeron-Pomeron processes
depends on the ion species. Roughly it scales with $Z^4/A^{1/3}$, see
\cite{Felix97}. Thus for heavy ions, like Pb, we may expect dominance
of the photon-photon processes whereas, say in $pp$-collisions, the
Pomeron-Pomeron processes will definitely dominate in coherent
collisions. 

Nevertheless diffractive events are of interest in ion collisions
too. As one is triggering again on an intact nucleus, one expects that
the coherent Pomeron emission from the whole nucleus will lead to a
total transverse momentum of the produced system similar to the
$\GG$-events. Therefore one expects that part of the events are coming
from diffractive processes. It is of interest to study how these could
be further distinguished from the photon-photon events.

Another class of background events are additional electromagnetic
processes. One of the dominating events here is the electromagnetic
excitation of the ions due to an additional single-photon exchange. As
mentioned above this is one of the dominant beam-loss processes for
Pb-Pb collisions. The probability to excite at least one of the ions
for Pb-Pb collisions is about 65\% and about 2\% for Ca-Ca for an impact
parameter of $2R$. Especially at large invariant masses, $\GG$-events
occur at impact parameter close to $2R$, therefore in the case of Pb-Pb
collisions one has to expect that most of them are accompanied by the
excitation or dissociation of one of the ions \cite{HenckenTB95,BaltzS98}. Most
of the excitation lead into the giant dipole resonance (GDR), which
has almost all of the dipole strength. As it decays predominantly via
the emission of a neutron, this leads to a relativistic neutron with
an energy of about 3 TeV in the forward direction. Similarly all other
low energy breakup reactions in the rest frame of one of the ions are
boosted to high energy particles in the laboratory. In order to
increase the $\GG$-luminosity it would be interesting to include these
events also in the $\GG$-trigger.  On the other hand one has
to make sure, that this does not obscure the interpretation of these
events as photon-photon events.

\begin{figure}[tbhp]
\begin{center}
\resizebox{7.5cm}{!}{\includegraphics{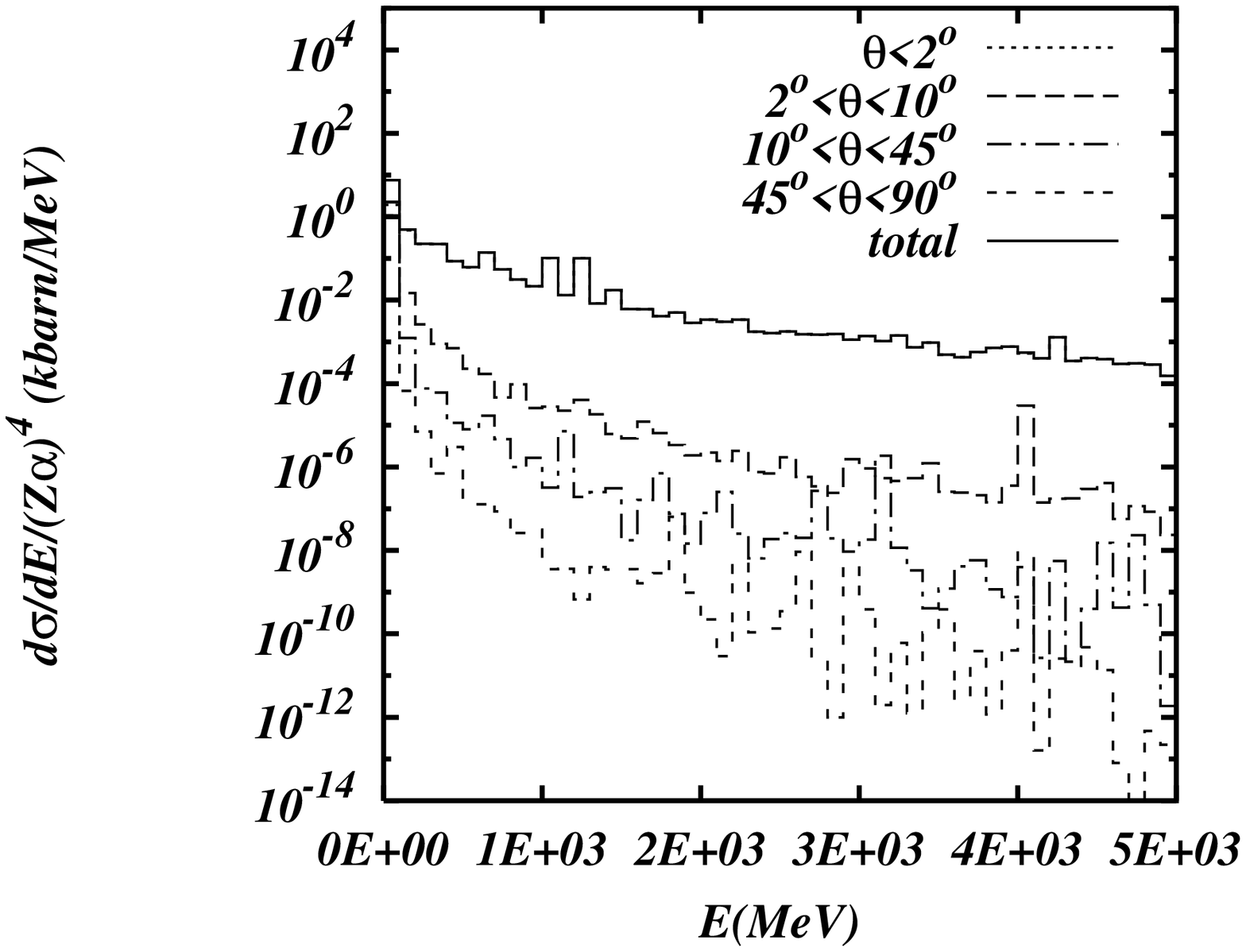}}
\resizebox{7.5cm}{!}{\includegraphics{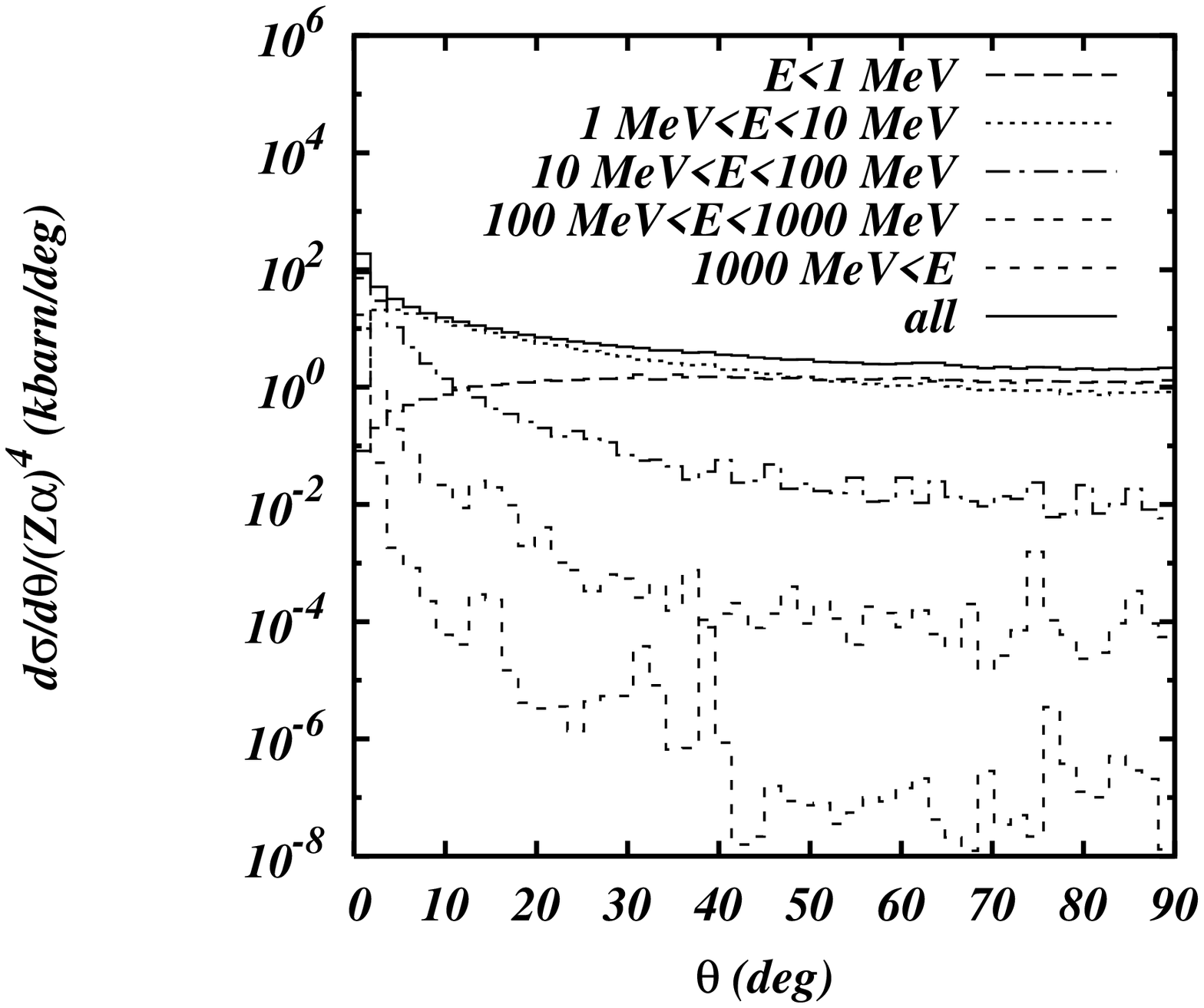}}
\end{center}
\caption{\it
The single differential cross section for a number of experimental
constraints. (a) for different angular-ranges as a function of energy,
(b) for different energies as a function of the angle
with the beam axis $\theta$.
}
\label{fig_eedifT}
\end{figure}

Another background process is the production of electron-positron
pairs, see Sec.~\ref{ssec_leptons}. Due to their small mass, they are
produced rather copiously. They are predominantly produced at low
invariant masses and energies and in the forward and backward
direction. Figure~\ref{fig_eedifT} shows cross section as a function
of energy and angle for different experimental cuts. 
On the other hand, as the total cross section for
this process is enormous ($\approx$ 230 kbarn for Pb-Pb collisions, 800
barn for Ca-Ca collisions), a significant cross section remains even at
high energies in the forward direction. This has to be taken into
account when designing forward detectors. Table~\ref{tab_ee} shows the
cross section for $e^+e^-$ production where the energy of both
particles is above a certain threshold value.
%
%
\begin{table}[htb]
\caption{
Cross sections of $e^+e^-$ pair production when {\em both} electron
and positron have an energy above a threshold value.}
\begin{center}
\begin{tabular}{c|rr}
\hline
$E_{thr}$ (GeV) & $\sigma$(Pb-Pb) & $\sigma$(Ca-Ca)\\
\hline
0.25& 3.5  kbarn & 12  barn \\
0.50& 1.5  kbarn & 5.5 barn \\
1.0 & 0.5  kbarn & 1.8 barn \\
2.5 & 0.08 kbarn & 0.3 barn \\
5.0 & 0.03 kbarn & 0.1 barn \\
\hline
\end{tabular}
\end{center}
\label{tab_ee}
\end{table}

\subsection{Conclusion}
\label{ssec_conclusion}

In this contribution to the CMS heavy ion chapter the basic properties
of peripheral hadron-hadron collisions are described. Electromagnetic
processes, that is, photon-photon and photon-hadron collisions, are an
interesting option, complementing the program for central collisions.
It is the study of events, with relatively small multiplicities and a 
small background. These are good conditions to search for new physics.
The method of equivalent photons is a well established tool to
describe these kinds of reactions. Reliable results of quasireal
photon fluxes and $\GG$-luminosities are available. Unlike electrons
and positrons heavy ions and protons are particles with an internal
structure. Effects arising from this structure are well under
control and minor uncertainties coming from the exclusion of central
collisions and triggering can be eliminated by using a luminosity
monitor from muon-- or electron--pairs. A trigger for peripheral
collisions is essential in order to select photon-photon events. Such
a trigger seems to be possible based on the survival of the nuclei
after the collision and the use of the small transverse momenta of the
produced system. A problem, which is difficult to judge quantitatively
at the moment, is the influence of strong interactions in grazing
collisions, i.e., effects arising from the nuclear stratosphere and
Pomeron interactions. 

The high photon fluxes open up possibilities for photon-photon as well
as photon-nucleus interaction studies up to energies hitherto
unexplored at the forthcoming colliders RHIC and LHC.  Interesting
physics can be explored at the high invariant $\GG$-masses, where
detecting new particles could be within range. Also very interesting
studies within the standard model, i.e., mainly QCD studies will be
possible. This ranges from the study of the total $\GG$-cross section
into hadronic final states up to invariant masses of about 100~GeV to
the spectroscopy of light and heavy mesons. The production via
photon-photon fusion complements the production from single photons in
$e^+$--$e^-$ collider and also in hadronic collisions via other
partonic processes.

Peripheral collisions using Photon-Pomeron and Pomeron-Pomeron
collisions, that is, diffractive processes are an additional
application. They use essentially the same triggering conditions and
therefore one should be able to record them at the same time as
photon-photon events.


\end{document}